\documentclass[12pt, draftclsnofoot, onecolumn]{IEEEtran}
\usepackage{bbm}
\usepackage{caption}
\usepackage{subfigure}
\usepackage{graphicx}
\usepackage{latexsym}
\usepackage{diagbox}
\usepackage{changepage}
\usepackage[fleqn]{amsmath}
\usepackage{amsmath}
\usepackage{amsfonts}
\usepackage{indentfirst}
\usepackage{CJK}
\usepackage{indentfirst}
\usepackage[varg]{txfonts}
\usepackage{stfloats}
\usepackage{multirow}%
\usepackage{booktabs}
\usepackage{color,soul}
\usepackage{epstopdf}
\usepackage{float}
\usepackage{bm}
\usepackage{makecell}
\usepackage{array}
\usepackage{bm}
\usepackage{mathtools}
\usepackage{geometry}
\usepackage[linesnumbered,ruled,commentsnumbered,longend]{algorithm2e}
\usepackage[linesnumbered,ruled,commentsnumbered,longend]{algorithm2e}
\makeatletter

\newcommand{\Rmnum}[1]{\expandafter\@slowromancap\romannumeral #1@}

\makeatother
\makeatletter

\newcommand{\qed}{\nobreak \ifvmode \relax \else
	\ifdim\lastskip<1.5em \hskip-\lastskip
	\hskip1.5em plus0em minus0.5em \fi \nobreak
	\vrule height0.75em width0.5em depth0.25em\fi}

\SetAlgoLongEnd

\ifCLASSINFOpdf
\else
\fi

\begin{document}

\title{\LARGE{Beamforming Analysis and Design for Wideband THz Reconfigurable Intelligent Surface Communications}}

\author{Wencai Yan, Wanming Hao,~\IEEEmembership{Member,~IEEE,} Chongwen Huang,~\IEEEmembership{Member,~IEEE,} Gangcan Sun, Osamu Muta,~\IEEEmembership{Member,~IEEE,}  Haris Gacanin,~\IEEEmembership{Fellow,~IEEE, } and Chau Yuen,~\IEEEmembership{Fellow,~IEEE}
	
	\thanks{W. Yan is with the School of Electrical and Information Engineering, Zhengzhou University, Zhengzhou 450001, China. (E-mail: yanwencai001@163.com)}
	\thanks{W. Hao is with the School of Electrical and Information Engineering, Zhengzhou University, Zhengzhou 450001, China, and also with the SongShan Laboratory, Zhengzhou 450018, China. (e-mail: iewmhao@zzu.edu.cn)}
	\thanks{G. Sun is with the School of Electrical and Information Engineering, Zhengzhou University, Zhengzhou 450001, China. (E-mail:iegcsun@zzu.edu.cn)}
	\thanks{C. Huang is with College of Information Science and Electronic Engineering, Zhejiang University, Hangzhou 310027, China, with the State Key Laboratory of Integrated Service Networks, Xidian University, Xi'an 710071, China, and with Zhejiang-Singapore Innovation and AI Joint Research Lab and Zhejiang Provincial Key Laboratory of Info. Proc., Commun. \& Netw. (IPCAN), Hangzhou 310027, China. (E-mail: chongwenhuang@zju.edu.cn ).}
	\thanks{O. Muta is with the Center for Japan-Egypt Cooperation in Science and Technology, Kyushu University, Fukuoka 819-0395, Japan.  (E-mails: muta@m.ieice.org)}
	\thanks{H. Gacanin is with the Institute for Communication Technologies and Embedded Systems, RWTH Aachen University, Aachen 52062, Germany. (E-mail:harisg@ieee.org)}
	\thanks{C. Yuen is with the School of Electrical and Electronics Engineering, Nanyang Technological University, 639798 Singapore (e-mail: chau.yuen@ntu.edu.sg)}}
\maketitle
\begin{abstract}
Reconfigurable intelligent surface (RIS)-aided terahertz (THz) communications have been regarded as a promising candidate for future 6G networks because of  its ultra-wide bandwidth and ultra-low power consumption. However, there exists the beam split problem, especially when the base station (BS) or RIS owns the large-scale antennas, which may lead to serious array gain loss. Therefore, in this paper, we investigate the beam split and beamforming design problems in the THz RIS communications. Specifically, we first analyze the beam split effect caused by different RIS sizes, shapes and deployments. On this basis, we apply the fully connected time delayer phase shifter hybrid beamforming (FC-TD-PS-HB) architecture at the BS and deploy distributed RISs to cooperatively mitigate the beam split effect. We aim to maximize the achievable sum rate by jointly optimizing the hybrid analog/digital beamforming, time delays at the BS and reflection coefficients at the RISs. To solve the formulated problem, we first design the analog beamforming and time delays based on different RISs' physical directions, and then it is transformed into an optimization problem by jointly optimizing the digital beamforming and reflection coefficients. Next, we propose an alternatively iterative optimization algorithm to deal with it. Specifically, for given the reflection coefficients, we propose an iterative algorithm based on the minimum mean square error technique to obtain the digital beamforming.
After, we apply Lagrangian dual reformulation (LDR) and multidimensional complex quadratic transform (MCQT) methods to transform the original problem to a quadratically constrained quadratic program, which can be solved by alternating direction method of multipliers (ADMM) technique to obtain the reflection coefficients.
Finally, the digital beamforming and reflection coefficients are obtained via repeating the above processes until convergence.
Simulation results verify that the proposed scheme can effectively alleviate the beam split effect and improve the system capacity.

\end{abstract}


%
\IEEEpeerreviewmaketitle

\section{Introduction}
To satisfy the requirement of the rapid growth of wireless data rates, terahertz (THz, 0.1-10 THz) with tens of GHz bandwidth are considered as one of the promising technologies for future 6G networks~\cite{refA}-\cite{ref70}. Nevertheless, THz signals usually suffer from severe attenuation and poor diffraction, which leads to the small coverage and short transmission distance. To address these issues, massive multiple-input multiple-output (mMIMO) technique can be applied to generate high-gain directional beams. However, the power consumption for conventional fully-digital structure is large due to the numerous radio frequency (RF) chains~\cite{refD}. Fortunately, several hybrid analog/digital beamforming structures are developed~\cite{refE},~\cite{refF}, where all antennas are connected to a reduced number of RF chains via phase shifters (PSs). In this way, the beamforming system is decomposed into a low-dimensional digital beamforming and a high-dimensional analog beamforming. Consequently, the required number of RF chains is effectively reduced, which significantly decreases the  power consumption~\cite{refE}. Furthermore, a large system capacity can still be obtained by optimizing the hybrid beamforming~\cite{refF}. Additionally, the poor diffraction problem still makes THz signals vulnerable to the obstruction. This results in a weak signal reception when there is no line-of sight (LoS) link. To solve this issue, reconfigurable intelligent surface (RIS) comprised of a large number of low-power passive elements can be deployed to generate a extra virtual LoS link between the base station (BS) and users. By adjusting the reflection coefficients of the RIS, the signal reception can be improved leading to the overall system-level performance improvement~\cite{refG},~\cite{ref100}.

Generally, the multiple carriers transmission techniques (e.g., orthogonal frequency division multiplexing, OFDM) are usually applied to overcome the frequency selective fading, especially for the ultra-wide bandwidth THz communications. However, a beam split problem, namely the beams of different subcarriers toward to different directions, results in the serious array gain loss for the hybrid analog/digital beamforing structure~\cite{ref13}. This is due to the frequency-independent property of PSs, where the phases controlled by PSs are common for all subcarriers. Furthermore, there may also exist beam split for the RIS. Therefore, overcoming this effect is significant challenge for the implementation of THz RIS communications. In this paper, we will focus on the above problem for the wideband THz RIS communications with beam split.
\subsection{Related Works}
Currently, there have been several works investigating the beam split problem. Initially,~\cite{refI} proposed a hybrid beamforming scheme and~\cite{refJ} designed a codebook-based beam selective scheme for reducing the array gain loss in the millimeter wave communication system. However, due to the much larger bandwidth of THz signals, the performance gain with above schemes in wideband THz communications is very limited.~\cite{ref30} proposed a double PSs-based hybrid analog/digital beamforming structure, and a hybrid analog and digital beamforming scheme was proposed to maximize the system capacity. Different from the conventional fully connected antenna structure, it needs more PSs for the proposed structure in~\cite{ref30} and leads to higher power consumption and hardware complexity.  Later, researchers propose to apply time delayers (TDs) between RF chains and PSs, effectively relieving the beam split effect because of their frequency-dependent property~\cite{ref25}-\cite{ref23}. For example,~\cite{ref25} proposed a dynamic subarray with fixed-true-time-delay structure, where the low-resolution discrete time delays are applied. The proposed scheme in~\cite{ref25} can effectively improve the system energy efficiency, but the beam split effect can not be solved well. Contrarily,~\cite{ref26} proposed a TD-based subarray hybrid beamforming structure, where each subarray is connected to a separate RF chain and each PS is connected to one TD. Furthermore, the time delay can be adjusted continuously. Whereas, the structure includes a lot of PSs, which results in the huge power consumption and complicated hardware design. Thus, \cite{ref23} proposed a TD-based hybrid beamforming architecture, where PSs are connected to RF chains only through a few TDs. After that, an effective scheme based on jointly optimizing time delay, analog and digital beamforming was proposed.

Although the above works considering the beam~split effect, the RIS and corresponding effect are not investigated. Nowadays, several works have started to study the RIS-based THz communications. In~\cite{ref31}, the authors aimed to maximize the achievable rate under a hybrid analog/digital beamforming architecture and proposed a deep learning-based multiple discrete classification hybrid beamforming scheme.~\cite{refK} proposed two effective hierarchical codebooks and beamforming design schemes to obtain a near-optimal performance.~\cite{ref101} investigated a novel hybrid beamforming architecture, where the digital beamforming matrix at the BS and analog beamforming  matrices at the RISs, for the multi-hop RIS-aided networks at THz-band frequencies. In addition, to maximize the weighted achievable rate by jointly optimizing the hybrid beamforming at the BS and reflection coefficients at the RIS,~\cite{refM} proposed an alternatively iterative optimization algorithm. For characterizing the capacity of the RIS-enabled THz MIMO system, an adaptive gradient descent method was proposed via dynamically updating the step size at each iteration~\cite{refN}.
However,~\cite{ref31}, \cite{refK}, \cite{ref101} only consider the single subcarrier. Although~\cite{refM}, \cite{refN} consider the wideband multiple subcarriers, the beam split problem is not investigated. Therefore, there has not been the related work jointly considering the beam split effect and beamforming design problems in wideband THz RIS communications. Besides, several practical RIS reflection models have been derived to obtain the phase-amplitude-frequency relationship of the reflected signals. For example, the authors of~\cite{addA}-\cite{addC} derived that the practical RIS reflection model is composed of complex arc-tangent function and Witch of Agnesi function. So far, the obtained practical RIS reflection models are based on low frequency.  Thus, for the THz frequency, we first study the RIS beam split with the ideal reflection model and obtain an upper bound of the system performance, similar to THz RIS research works \cite{refK}, \cite{ref101}, \cite{addD}.

\subsection{Main Contributions}
In this paper, we firstly analyze the beam split effect for the wideband THz RIS communications. To alleviate this effect, we then investigate the joint beamforming design problem. The main contributions are summarized as follows:
\begin{itemize}
\item[$\bullet$]
We construct the wideband THz RIS channel model and analyze the beam split effect. Specifically, we firstly derive the achievable array gain at different subcarriers. To understand the system behaviour, we analyze and discuss the array gain loss of the system. Finally, to address and understand the system behaviour under various configurations, we study the beam split effect under different  RIS sizes, shapes and deployments by theoretical analysis and simulations.
\end{itemize}
\begin{itemize}
\item[$\bullet$]
To reduce the beam split effect, we introduce the fully connected time delay phase shifter for hybrid beamforming (FC-TD-PS-HB) architecture at the BS and distributed RISs deployment. On this basis, we formulate a sum rate maximization problem via jointly optimizing the hybrid analog/digital beamforming, time delays at the BS and reflection coefficients at the RISs. Due to high complexity of the formulated problem, we first design the analog beamforming and time delays according to the different RISs' physical directions, and then the original problem is converted to an optimization problem by jointly optimizing the digital beamforming and reflection coefficients.
\end{itemize}
\begin{itemize}
\item[$\bullet$]
We propose an alternatively iterative optimization scheme. Specifically, we firstly fix the reflection coefficients, based on the equivalence between the sum rate  maximization and sum mean square error (MSE) minimization, we propose a  minimum mean square error (MMSE)-based iterative algorithm for digital beamforming. Next, we utilize Lagrangian dual reformulation (LDR) to decouple the logarithms and apply multidimensional complex quadratic transform (MCQT) method to address the non-convexity problem of the high-dimensional fractions. After that, the original problem is transformed in a quadratically constrained quadratic program (QCQP) problem, which is solved by alternating direction method of multipliers (ADMM) technique. Finally, the above procedures are repeated until convergence, and the final solutions are obtained.

\end{itemize}

Notations: Lower-case and upper-case boldface letters represent vectors and matrices, respectively. $(\cdot)^{T},(\cdot)^{H}$ denote the transpose and Hermitian transpose, respectively. $|\cdot|$ denotes the absolute operator. $\left\|\cdot \right\|$ is the Frobenius norm. $\mathbf{I}_{N}$ denotes the identity matrix of size $N \times N$. $\rm diag(\cdot)$ represents diagonal operation. $\mathbb{C}^{x \times y}$ denotes the space of $x \times y$ complex matrix.  $\rm {Re}(\cdot)$ means real number operation, $\mathcal{C} \mathcal{N}\left(A, B \right)$ represents the Gaussian distribution with mean $A$ and covariance $B$.
\begin{figure*}[t]
	\centering
	\subfigure[]{
		\label{Iteration} 
		\includegraphics[width=7cm,height=4cm]{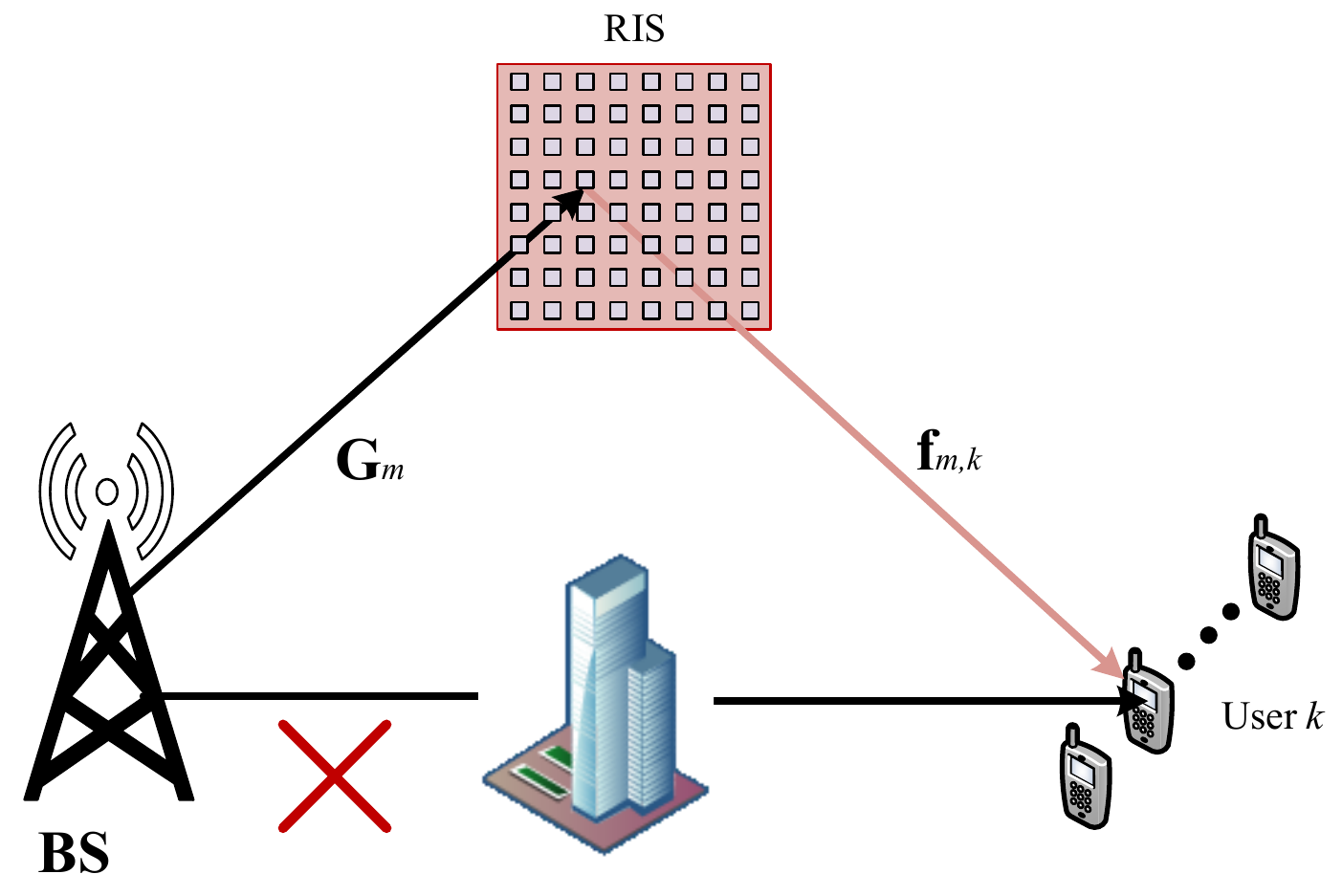}}
	\subfigure[]{
		\label{complexity} 
		\includegraphics[width=7cm,height=4cm]{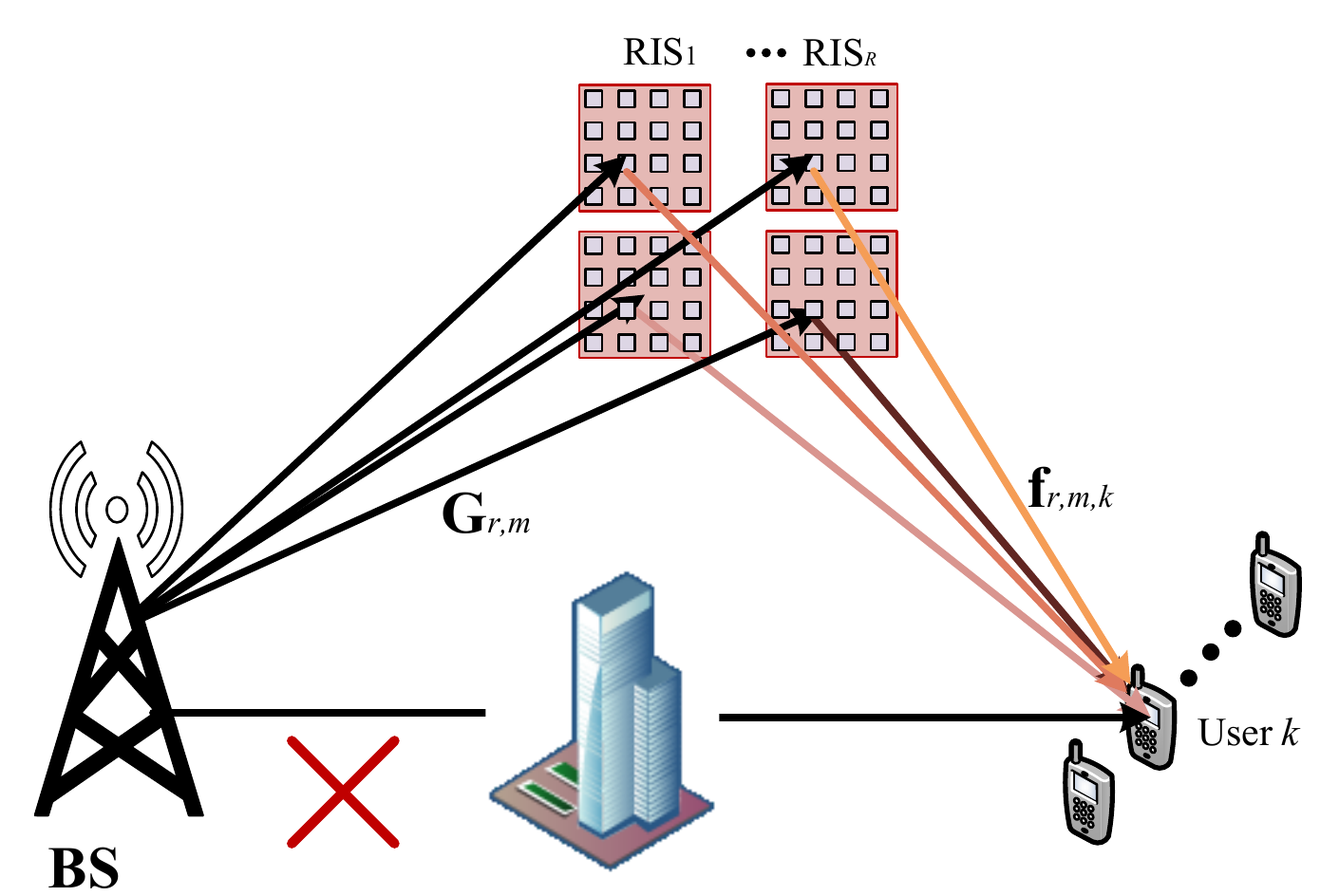}}
	\caption{(a) The system model for wideband THz centralized RIS communications. (b) The system model for wideband THz distributed RIS communications.}
\end{figure*}
\section{Channel Model and Beam Split Effect Analysis}
In this section, we first give the channel model for wideband THz RIS communications, and then analyze the beam split effect at the BS and RIS.
\subsection{Channel Model}
We investigate a wideband THz RIS communication system with hybrid beamforming architecture at the BS\footnote{The fully connected and sub-array hybrid structures at the BS are two classic antenna structures. Generally, the design and beamforming optimization of the fully connected hybrid structure are more complex and challenging than that of the sub-array hybrid structure. Therefore, we mainly consider the fully connected hybrid structure, and the corresponding design and analysis can be directly applied to the sub-array hybrid structure.}. The BS is equipped with the uniform linear array with $N_{{\rm{TX}}}$ antennas and  $N_{{\rm{RF}}}$ $(N_{{\rm{TX}}} \geq N_{{\rm{RF}}})$ RF chains to serve $K$ single-antenna uesrs. The RIS is consisted of a uniform planar array with $N_{\rm {RIS}}={M_{x}}\times{M_{y}}$ elements, where $M_{x}$ and $M_{y}$ represent the number of rows and columns, respectively. We define $\mathcal{M}_{x}=\{1, \cdots, M_{x}\}$, $\mathcal{M}_{y}=\{1, \cdots, M_{y}\}$ as the index sets of elements on rows and columns~\cite{ref5}. In fact, there are already some existing works that proposed many effective channel estimation algorithms for RIS-based system, e.g., \cite{ref102} proposed a parallel factor analysis (PARAFAC)-based method. Since channel estimation is not main focus of this work, then, we assume that the channel information is obtained by existing channel estimation algorithms. Additionally, we assume that the direct links from the BS to users are blocked by obstacles, as shown in Fig. 1. Consequently, the equivalent channel $\mathbf{h}_{m, k}$ between the BS and the $k$-th user on the $m$-th subcarrier can be represented as
\begin{eqnarray}
  \mathbf{h}_{m, k}= \mathbf{f}_{ m, k} \mathbf{\Phi} \mathbf{G}_{m},
\end{eqnarray}
where $\mathbf{f}_{m, k} \in \mathbb{C}^{1 \times N_{\rm{RIS}}}$ denotes the channel between the RIS and the $k$-th user on the $m$-th subcarrier, $\mathbf{G}_{m} \in \mathbb{C}^{N_{\rm {RIS}} \times N_{\rm {TX}}}$ represents the channel from the BS to RIS on the $m$-th subcarrier. $\mathbf{\Phi}=\operatorname{diag}\left(\varphi_{1,1}, \cdots, \varphi_{m_{x},m_{y}}, \cdots, \varphi_{M_{x},M_{y}}\right)$ is the RIS reflection coefficients matrix, where $\varphi_{m_{x},m_{y}}= \varepsilon_{m_{x},m_{y}} e^{j \phi_{m_{x},m_{y}}}$ with $m_{x} \in \mathcal{M}_{x}$, $m_{y} \in \mathcal{M}_{y}$ and $\varepsilon_{m_{x}, m_{y}} \in[0,1]$, $\phi_{m_{x},m_{y}} \in[0, 2 \pi)$ respectively represent the reflection amplitude and phase. We define $\boldsymbol{\varphi}=\left[\varphi_{1,1}, \cdots, \varphi_{m_{x},m_{y}}, \cdots, \varphi_{M_{x},M_{y}}\right]^{ T}$ as the reflection coefficients vector of the RIS.

We apply the Saleh-Valenzuela THz channel model~\cite{ref2}. The frequency at the $m$-th subcarrier is $f_{m}=f_{c}+\frac{B}{M}\left(m-1-\frac{M-1}{2}\right), m=1,2,\cdots, M$, where $f_{c}$ and $B$ are the central frequency and bandwidth, respectively. Therefore, the frequency-domain channel matrix $\mathbf{G}_{m}$ of the BS-RIS link can be expressed as
\begin{eqnarray}
\mathbf{G}_{m}=\sum_{l_{1}=1}^{L_{1}} \alpha_{l_{1}} e^{-j 2 \pi \tau_{l_{1}} f_{m}} \mathbf{b}\left(u_{l_{1}}, v_{l_{1}}\right) \mathbf{a}\left(\theta_{l_{1}}\right)^{ H},
\end{eqnarray}
where $L_{1}$ represents the number of paths, $\alpha_{l_{1}}$ and $\tau_{l_{1}}$ respectively denote the gain and delay of the $l_{1}$-th path, $\mathbf{a}\left(\theta_{l_{1}}\right)$ and $\mathbf{b}\left(u_{l_{1}}, v_{l_{1}}\right)$ respectively denote the array steering vectors at the BS and RIS, which can be denoted as
\begin{eqnarray}
 \mathbf{a}\left(\theta_{l_{1}}\right)
=\frac{1}{\sqrt{N_{\rm {TX}}}}\left[1, \ldots, e^{j 2 \pi d \frac{f_{m}}{c} n_{\rm{TX}} \sin \theta_{l_{1}}}, \ldots, e^{\left.j 2 \pi d \frac{f_{m}}{c}\left(N_{\rm {TX}}-1\right) \sin \theta_{l_{1}}\right)}\right]^{ T},
\end{eqnarray}
\begin{eqnarray}
\begin{split}
\mathbf{b}\left(u_{l_{1}}, v_{l_{1}}\right)
=\frac{1}{\sqrt{N_{\rm{RIS}}}}[1, \ldots, e^{j 2 \pi d \frac{f_{m}}{c}(m_{x} \sin  u_{l_{1}} \sin v_{l_{1}}+m_{y} \cos v_{l_{1}})},
\ldots, e^{j 2 \pi d \frac{f_{m}}{c}((M_{x}-1) \sin  u_{l_{1}} \sin v_{l_{1}}+(M_{y}-1) \cos v_{l_{1}})}]^{ T}.
\end{split}
\end{eqnarray}
Here, we assume the elements of RIS lying on the y and z-axes~\cite{add3}, $d$ denotes the element spacing and is usually set as $d=\lambda_{c} / 2$, where $\lambda_{c}$ means the wavelength of the central frequency $f_{c}$.  $\theta_{l_{1}} \in[-\pi / 2, \pi / 2]$ is the physical direction of the $l_{1}$-th path departing from the BS, and $\eta_{l_{1},m}= \frac{f_{m}}{f_c} \sin \theta_{l_{1}}$ is denoted as the spatial direction. $u_{l_{1}}\in[-\pi / 2, \pi / 2]$ and $v_{l_{1}} \in[0, \pi]$ represent the azimuth and elevation angles of arrivals (AOAs) at the RIS for the $l_{1}$-th path, respectively. Next, the frequency-domain channel vector between the RIS and the $k$-th user can be expressed as
\begin{eqnarray}
\mathbf{f}_{m, k}=\sum_{l_{2}=1}^{L_{2}} \alpha_{r, l_{2}} e^{-j 2 \pi \tau_{l_{2}} f_{m}} \mathbf{b}\left(u_{l_{2}}^{k}, v_{l_{2}}^{k}\right)^{T},
\end{eqnarray}
where $L_{2}$ represents the number of paths, $\alpha_{l_{2}}$ and $\tau_{l_{2}}$ respectively denote the gain and delay of the $l_{2}$-th path, $u_{l_{2}}^{k}\in[-\pi / 2, \pi / 2]$ and $v_{l_{2}}^{k} \in[0, \pi]$ represent the azimuth and elevation angles of departures (AODs) from the RIS for the $l_{2}$-th path, respectively.
\subsection{Beam Split Effect Analysis}
In this subsection, we analyze the beam split effect at the BS and RIS. In fact, there have been several works investigating the beam split effect at the BS, such as literature~\cite{ref25}-\cite{ref23}. The main reasons are the ultra-wide bandwidth of the THz signals and the large-scale antennas at the BS. When the hybrid beamforming structure at the BS is applied, the frequency-independent property of PSs leads to the array gain loss~\cite{ref23}. Meanwhile, the RIS may also exist beam split. Here, we mainly analyze the beam split effect at the RIS, and that at the BS can refer to~\cite{ref25}-\cite{ref23}.

We first analyze the beam split effect under different RIS sizes (elements), and then investigate that under different RIS shapes and deployments. For the sake of analysis, we consider the single-antenna BS and single user. Let $\mathbf{g}(t)$ and $\mathbf{f}(t)$ denote the time-domain channel vectors of the BS-RIS link and RIS-user link, respectively. Then, we denote $\tau_{l_{1}, m_{x},m_{y}}$ as the time delay of the $l_{1}$-th path spanning from the BS to the $(m_{x},m_{y})$-th element of the RIS and $\tau_{l_{2}, m_{x},m_{y}}$ as the time delay of the $l_{2}$-th path spanning from the $(m_{x},m_{y})$-th element of the RIS to the user. Based on the far-field assumption that the RIS planar size is much smaller than the distance between the transmitter and receiver, the path delay $\tau_{l_{1}, m_{x},m_{y}}$ and $\tau_{l_{2}, m_{x},m_{y}}$ can be written~as
\begin{eqnarray}\label{ZH}
\begin{aligned}
\tau_{l_{1}, m_{x},m_{y}}&=\tau_{l_{1}}+\frac{d\left(m_{x}-1\right) \sin  u_{l_{1}} \sin  v_{l_{1}}+d\left(m_{y}-1\right) \cos v_{l_{1}}}{c}\\ &=\tau_{l_{1}}+\frac{\left(m_{x}-1\right) \bar{u}_{l_{1}}+\left(m_{y}-1\right) \bar{v}_{l_{l}}}{f_{c}},
\end{aligned}
\end{eqnarray}

\begin{eqnarray}\label{ZI}
\begin{aligned}
\tau_{l_{2}, m_{x},m_{y}}&=\tau_{l_{2}}-\frac{d\left(m_{x}-1\right) \sin u_{l_{2}} \sin  v_{l_{2}}+d\left(m_{y}-1\right) \cos v_{l_{2}}}{c}\\ &=\tau_{l_{2}}-\frac{\left(m_{x}-1\right) \bar{u}_{l_{2}}+\left(m_{y}-1\right) \bar{v}_{l_{2}}}{f_{c}},
\end{aligned}
\end{eqnarray}
where $m_{x} \in \mathcal{M}_{x}$, $m_{y} \in \mathcal{M}_{y}$, $\tau_{l_{1}} \triangleq \tau_{l_{1}, 1,1}$ and $\tau_{l_{2}} \triangleq \tau_{l_{2}, 1,1}$ for notational simplicity. We define $\bar {u}_{l_{1}}=\frac{\sin u_{l_{1}} \sin  v_{l_{1}}}{2}$ and $\bar{v}_{l_{1}}=\frac{ \cos v_{l_{1}}}{2}$ as the normalized elevational and azimuth AOAs at the RIS, respectively. $\bar{u}_{l_{2}}=\frac{ \sin u_{l_{2}}\sin v_{l_{2}}}{2}$ and $\bar{v}_{l_{2}}=\frac{ \cos v_{l_{2}}}{2}$, respectively, are denoted as the normalized elevational and azimuth AODs at the RIS.
Therefore, the impulse response of the channel from the BS to the $(m_{x},m_{y})$-th RIS element and the channel from the $(m_{x},m_{y})$-th RIS element to the user can be expressed as~\cite{ref4}
\begin{eqnarray}\label{ZD}
\begin{aligned}
g_{m_{x},m_{y}}(t)&=\sum_{l_{1}=1}^{L_{1}} \alpha_{l_{1}} e^{-j 2 \pi f_{c} \tau_{l_{1}, m_{x},m_{y}}} \delta\left(t-\tau_{l_{1}, m_{x},m_{y}}\right),
\end{aligned}
\end{eqnarray}

\begin{eqnarray}\label{ZE}
\begin{aligned}
f_{m_{x},m_{y}}(t)&=\sum_{l_{2}=1}^{L_{2}} \alpha_{l_{2}} e^{-j 2 \pi f_{c} \tau_{l_{2}, m_{x},m_{y}}} \delta\left(t-\tau_{l_{2}, m_{x},m_{y}}\right).
\end{aligned}
\end{eqnarray}
The received reflecting signal via the $(m_{x},m_{y})$-th RIS element at the user can be represented as
\begin{eqnarray}\label{ZF}
y_{m_{x},m_{y}}(t) &= f_{m_{x},m_{y}}(t) * \varphi_{m_{x},m_{y}} g_{m_{x},m_{y}}(t)*s(t)+n(t).
\end{eqnarray}
Substituting (\ref{ZD}) and (\ref{ZE}) into (\ref{ZF}), we have
\begin{eqnarray}
\begin{aligned}y_{m_{x},m_{y}}(t)&=\varphi_{m_{x},m_{y}} \sum_{l_{1}=1}^{L_{1}} \sum_{l_{2}=1}^{L_{2}} \alpha_{l_{1}} \alpha_{l_{2}} e^{-j 2 \pi f_{c} \tau_{l_{1}, m_{x},m_{y}}} e^{-j 2 \pi f_{c} \tau_{l_{2}, m_{x},m_{y}}}
s\left(t-\tau_{l_{1}, m_{x},m_{y}}-\tau_{l_{2},m_{x},m_{y}}\right)+n(t)
\\ &=\varphi_{m_{x},m_{y}} \bar h_{m_{x},m_{y}}(t) * s(t)+n(t), \end{aligned}
\end{eqnarray}
where $\bar h_{m_{x},m_{y}}(t)$ is the $(m_{x},m_{y})$-th cascaded BS-RIS-user element channel impulse response and can be represented as
\begin{eqnarray}\label{ZG}
\begin{aligned}
\bar h_{m_{x},m_{y}}(t)&=\sum_{l_{1}=1}^{L_{1}} \sum_{l_{2}=1}^{L_{2}} \alpha_{l_{1}} \alpha_{l_{2}} e^{-j 2 \pi f_{c} \tau_{l_{1}, m_{x},m_{y}}} e^{-j 2 \pi f_{c} \tau_{l_{2}, m_{x},m_{y}}}\delta\left(t-\tau_{l_{1}, m_{x},m_{y}}-\tau_{l_{2},m_{x},m_{y}}\right).
\end{aligned}
\end{eqnarray}
Taking the Fourier transform to (\ref{ZG}), the frequency response of the $(m_{x},m_{y})$-th cascaded BS-RIS-user element channel can be denoted as
\begin{eqnarray}\label{ZJ}
\begin{aligned}
\bar h_{m_{x},m_{y}}(f)&=\sum_{l_{1}=1}^{L_{1}} \sum_{l_{2}=1}^{L_{2}} \alpha_{l_{1}} \alpha_{l_{2}} e^{-j 2 \pi f_{c} \tau_{l_{1}, m_{x},m_{y}}} e^{-j 2 \pi f_{c} \tau_{l_{2}, m_{x},m_{y}}}e^{-j 2 \pi f \left(\tau_{l_{1}, m_{x},m_{y}}+\tau_{l_{2},m_{x},m_{y}}\right)},
\end{aligned}
\end{eqnarray}
where $f \in [0,B]$. Then, substituting (\ref{ZH}) and (\ref{ZI}) into (\ref{ZJ}), we have
\begin{eqnarray}
\begin{aligned}
\bar h_{m_{x},m_{y}}(f)&= \sum_{l_{1}=1}^{L_{1}} \sum_{l_{2}=1}^{L_{2}} \bar{\alpha}_{l_{1}} \bar{\alpha}_{l_{2}} e^{-j 2\pi\left(m_{x}-1\right)\left(\bar u_{l_{1}}-\bar u_{l_{2}}\right)\left(1+\frac{f}{f_{c}}\right)}
e^{-j 2 \pi\left(m_{y}-1\right)\left(\bar v_{l_{1}}-\bar v_{l_{2} }\right)\left(1+\frac{f}{f_{c}}\right)} e^{-j 2 \pi f\left(\tau_{l_{1}}+\tau_{l_{2}}\right)}
\\&=\sum_{l_{3}=1}^{L_{1}L_{2}} c_{l_{3}} e^{-j 2 \pi\left(m_{x}-1\right)u_{l_{3}}\left(1+\frac{f}{f_{c}}\right)}e^{-j 2 \pi\left(m_{y}-1\right)v_{l_{3}}\left(1+\frac{f}{f_{c}}\right)} e^{-j 2 \pi f \tau_{l_{3}}}
\\&=\sum_{l_{3}=1}^{L_{1}L_{2}} c_{l_{3}} e^{-j 2 \pi\left(1+\frac{f}{f_{c}}\right)\left[\left(m_{x}-1\right)u_{l_{3}}+\left(m_{y}-1\right)v_{l_{3}}\right]} e^{-j 2 \pi f\tau_{l_{3}}},
\end{aligned}
\end{eqnarray}
where $\bar{\alpha}_{l_{1}}=\alpha_{l_{1}} e^{-j 2 \pi f_{c} \tau_{l_{1}}}$, $\bar{\alpha}_{l_{2}}=\alpha_{l_{2}} e^{-j 2 \pi f_{c} \tau_{l_{2}}}$, and we define $c_{l_{3}}=\bar{\alpha}_{l_{1}}\bar{\alpha}_{l_{2}}$, $v_{l_{3}}=\bar v_{l_{1}}-\bar v_{l_{2}}$, $u_{l_{3}}=\bar u_{l_{1}}-\bar u_{l_{2}}$, $\tau_{l_{3}}=\tau_{l_{1}}+\tau_{l_{2}}$, $l_{3} \in\left\{1,2, \cdots, L_{1} L_{2}\right\}$. Stacking $\bar h_{m_{x},m_{y}}(f)$ from the RIS elements into a vector yield, we have $\mathbf{\bar h}(f)=\left[\bar h_{1,1}(f), \cdots, \bar h_{m_{x},m_{y}}(f), \cdots, \bar h_{M_{x},M_{y}}(f)\right]^{\mathrm{T}}$, namely
\begin{eqnarray}
\mathbf{\bar h}(f)=\sum_{l_{3}=1}^{L_{1} L_{2}} c_{l_{3}} \mathbf{\bar b}\left(u_{l_{3}},v_{l_{3}}\right) e^{-j 2 \pi f \tau_{l_{3}}},
\end{eqnarray}
where
\begin{eqnarray}\label{ZK}
\begin{aligned}
\mathbf{\bar b}\left(u_{l_{3}},v_{l_{3}}\right)&=\left[ 1, \cdots, e^{-j 2 \pi \left(1+\frac{f}{f_{c}}\right) \left((m_{x}-1)u_{l_{3}}+(m_{y}-1)v_{l_{3}}\right)},
 \cdots, e^{-j 2 \pi \left(1+\frac{f}{f_{c}}\right) \left(\left(M_{x}-1\right)u_{l_{3}}+\left(M_{y}-1\right)v_{l_{3}}\right)}
\right]^{T},
\end{aligned}
\end{eqnarray}
is the spatial-domain steering vector. Different from the channel models widely used in~\cite{ref7}, the steering vectors in (\ref{ZK}) is frequency-dependent.We denote $u=2 u_{l_{3}}$, $v=2 v_{l_{3}}$ for simplicity, and assume $\varepsilon_{m_{x},m_{y}}=1$, $m_{x} \in \mathcal{M}_{x}$, $m_{y} \in \mathcal{M}_{y}$. In addition, due to the severe loss induced by the scattering, the THz communications heavily depend on the LoS path~\cite{ref50}, and thus we set $L_{1}=L_{2}=1$. Next, we define $\Gamma\left(f, u, v, \phi\right)$ as the array gain at arbitrary frequency $f$ with equivalent direction $(u,v)$, which can be written as
\begin{eqnarray}
\begin{aligned}
&\Gamma\left(f, u, v, \phi, M_{x}, M_{y}\right)=\left|\sum_{m_{x}=1}^{M_{x}} \sum_{m_{y}=1}^{M_{y}} e^{j\left\{\phi_{m_{x}, m_{y}}-\pi\left(1+\frac{f}{f_{c}}\right)\left[\left(m_{x}-1\right) u+\left(m_{y}-1\right) v\right]\right\}}\right|.
\end{aligned}
\end{eqnarray}
We assume that the equivalent direction $u=u_{0}$, $v=v_{0}$ for the central frequency, i.e., $f=f_{c}$. One can observe that the array gain $\Gamma\left(f, u_{0}, v_{0}, \phi, M_{x}, M_{y}\right)$ reaches its maximum value when $\phi_{m_{x}, m_{y}}-\pi\left(1+\frac{f_{c}}{f_{c}}\right)[\left(m_{x}-1\right) u_{0}\\+\left(m_{y}-1\right) v_{0}]=0$, namely $\bar\phi_{m_{x}, m_{y}}=2\pi[\left(m_{x}-1\right) u_{0}+\left(m_{y}-1\right) v_{0}]$. Therefore, the optimal reflection coefficients vector can be expressed as
\begin{eqnarray}
\boldsymbol{\bar \varphi}&=\left[ 1, \cdots, e^{j 2 \pi  \left[(m_{x}-1)u_{0}+(m_{y}-1)v_{0}\right]},
 \cdots, e^{j 2 \pi \left[\left(M_{x}-1\right)u_{0}+\left(M_{y}-1\right)v_{0}\right]}
\right]^{\mathrm{T}}.
\end{eqnarray}
Next, we give the array gain at arbitrary equivalent direction $(u,v)$ and frequency $f$ by setting $\boldsymbol{\varphi}=\boldsymbol{\bar \varphi}$, and have
\begin{eqnarray}
\begin{aligned}
\Gamma\left(f, u, v, \bar\phi,M_{x}, M_{y}\right)&=\left|\sum_{m_{x}=1}^{M_{x}} e^{j \pi\left\{\left(m_{x}-1\right)\left[2 u_{0}-\left(1+\frac{f}{f_{c}}\right) u\right]\right\}}
\sum_{m_{y}=1}^{M_{y}} e^{j \pi\left\{\left(m_{y}-1\right)\left[2 v_{0}-\left(1+\frac{f}{f_{c}}\right) v\right]\right\}}\right|.
\end{aligned}
\end{eqnarray}
It is easy to found that when $2 u_{0}-\left(1+\frac{f}{f_{c}}\right) u=0, 2 v_{0}-\left(1+\frac{f}{f_{c}}\right) v=0$ is satisfied, i.e., the equivalent direction $(u,v)=(\frac{2}{\left(1+\frac{f}{f_{c}}\right)} u_{0}, \frac{2}{\left(1+\frac{f}{f_{c}}\right)} v_{0})$, and the array gain reaches its maximum value. For a narrowband system, the signal frequency $f$ satisfies $f \approx f_{\mathrm{c}}$ across the entire bandwidth $B$, and thus the equivalent direction satisfies $(u,v)=(u_{0}, v_{0})$, i.e., the beams at all subcarriers toward to the same direction $(u_{0}, v_{0})$. Therefore, there is no array gain loss. However, for a wideband system, the signal frequency $f$ can not be approximated as $f_{c}$, which means that the beams reflected by the RIS may split into different physical directions over different subcarriers, which leads to the serious array gain loss.
\begin{figure}[t] 
	\centering
	\subfigure[]{
		\label{Iteration} 
		\includegraphics[width=7.5cm,height=5cm]{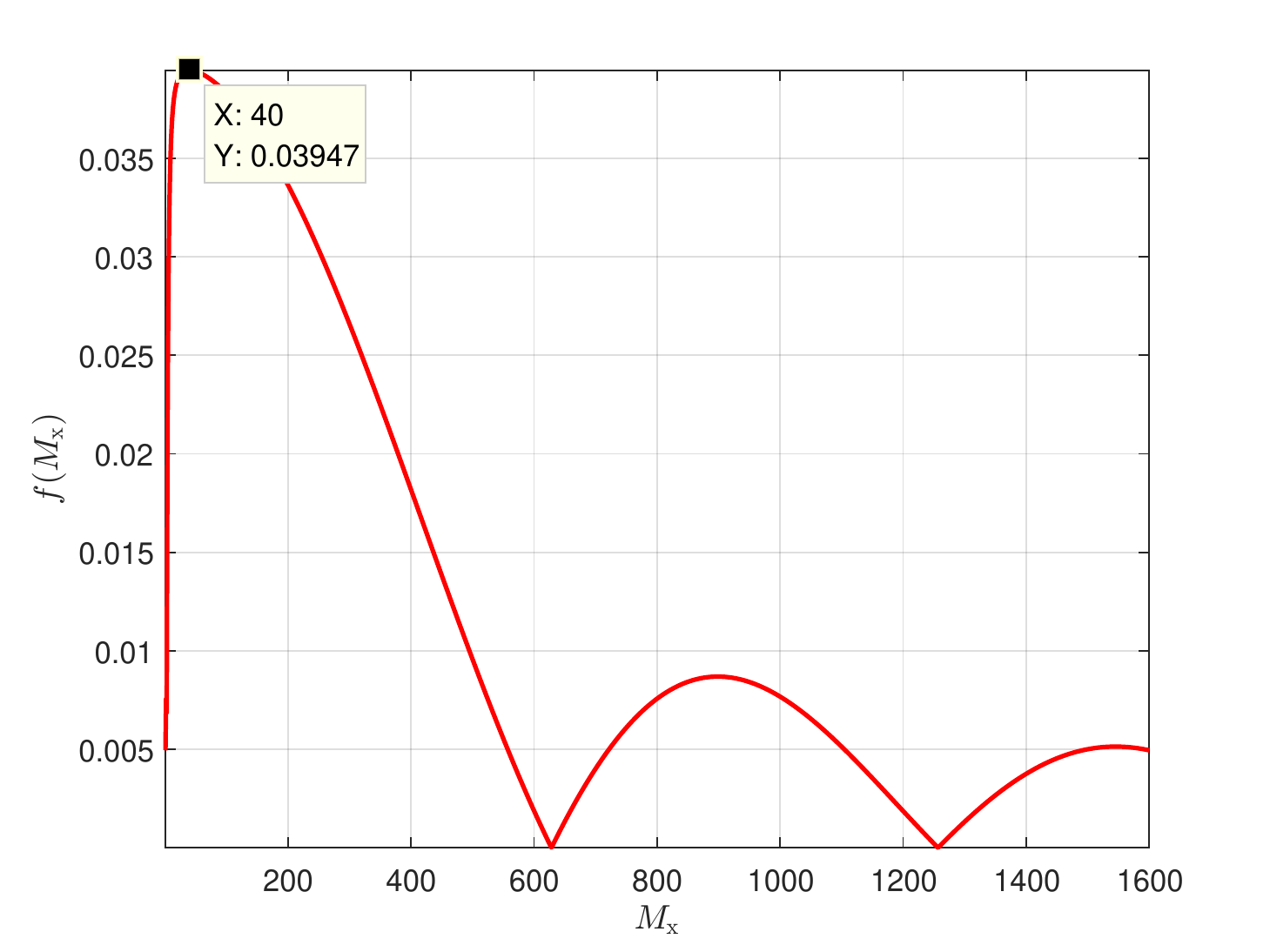}}
	\subfigure[]{
		\label{complexity} 
		\includegraphics[width=7.5cm,height=5cm]{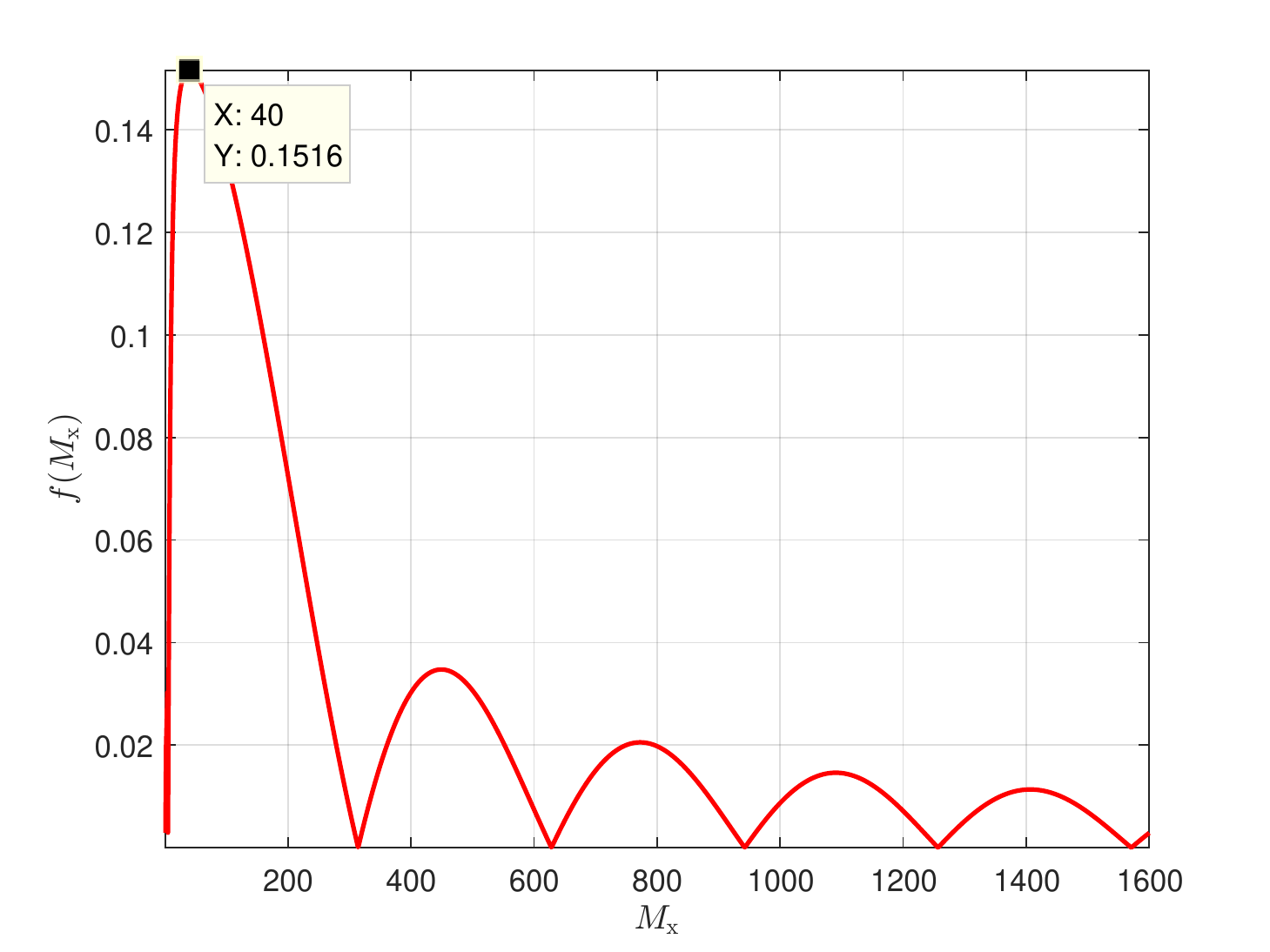}}
	\caption{The function $f(M_{x})$ against the number of rows $M_{\rm{x}}$. (a) $a=0.005$. (b) $a=0.01$.}
\end{figure}

Then, we analyze the beam split effect under different number RIS sizes and shapes. We first derive the array gain at equivalent direction $(u_{0}, v_{0})$ and arbitrary frequency $f$ by setting $\boldsymbol{\varphi}=\boldsymbol{\bar \varphi}$, which can be calculated as
\begin{eqnarray}\label{ZL}
\begin{aligned}
\Gamma\left(f, u_{0}, v_{0}, \bar\phi, M_{x}, M_{y}\right)&=\left|\sum_{m_{x}=1}^{M_{x}} e^{j \pi\left\{\left(m_{x}-1\right) \hat{u}\right\}}\sum_{m_{y}=1}^{M_{y}} e^{j \pi\left\{\left(m_{y}-1\right)\hat{v}\right\}}\right|\\
&=\left|\frac{\sin \left(\frac{\pi M_{x}}{2} \hat{u}\right)}{\sin \left(\frac{\pi}{2} \hat{u}\right)} \frac{\sin \left(\frac{\pi M_{y}}{2} \hat{v}\right)}{\sin \left(\frac{\pi}{2} \hat{v}\right)}\right|,
\end{aligned}
\end{eqnarray}
where $\hat{u}=\left(1-\frac{f}{f_{c}}\right) u_{0}$, $\hat{v}=\left(1-\frac{f}{f_{c}}\right) v_{0}$.
Based on the above analysis, we compare the beam split effect under different shapes with the total number of fixed RIS elements $M_{x}M_{y}$. Thus, the array gain can be expressed as
\begin{eqnarray}
\Gamma\left(M_{x}, M_{y}\right)=\left|\frac{\sin \left(a M_{x}\right)}{\sin a } \frac{\sin \left(b M_{y} \right)}{\sin b}\right|,
\end{eqnarray}
where $a=\frac{\pi \hat{u}}{2}$, $b=\frac{\pi \hat{v}}{2}$.
To make it more obvious that the relationship between array gain and RIS shapes, we first define
\begin{eqnarray}
f(M_{x},M_{y})=\left|\sin(a M_{x})\sin(b M_{y})\right|.
\end{eqnarray}
Since the total number of RIS elements $M_{x}M_{y}$ is fixed, we assume $M_{x}M_{y}=z$, and then (22) can be transformed as
\begin{eqnarray}
f(M_{x})=\left|\sin(a M_{x})\sin(b \frac{z}{M_{x}})\right|.
\end{eqnarray}
Since the $sin$ is a non-monotonic function, it is extremely difficult to directly give the optimal $M_x$ by theory proof when  $f(M_x)$ reaches its maximum value. Generally, it is obvious that $a$ and $b$ are very small for the THz communication, and we first assume that they are very small values with $a=b$. Next, we give the optimal $M_x$ by the following simulation. Fig. 2 depicts the function $f(M_{x})$ against the number of rows $M_{\rm{x}}$ when $a=0.005$ and $a=0.01$, and $z$ is set as 1600. From Fig. 2 (a) and (b), it can be observed that the function $f(M_{x})$ reaches its maximum value when $M_{x}=M_{y}=40$. These results indicate that the array gain loss under the square shape being smaller than that under the rectangular~shape.  Note that the conclusion is based on the above reasonable assumption. 

Moreover, Fig.~3 shows the normalized array gain under five square RISs with different elements, where $f_{c}=100$ GHz, $B=10$ GHz, $M=128$, $(u_{0},v_{0})=(0.5,0.5)$. One can observe that as the number of RIS elements increases, the array gain loss becomes larger for each subcarrier, which means that there exists serious beam split. The main reason is that more elements lead to larger signal delay, and thus bringing larger beam split according to~\cite{ref60}. Additionally, we show the normalized array gain under different RIS shapes based on the same number of elements as illustrated in Fig.~4.  Without loss of generality, we give all possible RIS shapes. From Fig.~4, we can find that the array gain loss under the square shape is smaller than that under the rectangular shape. Furthermore,  more flat shape leads to larger array gain loss.
\begin{figure}[htbp]
\centering
    \label{RIS_subcarrier} 
    \includegraphics[width=9cm,height=6cm]{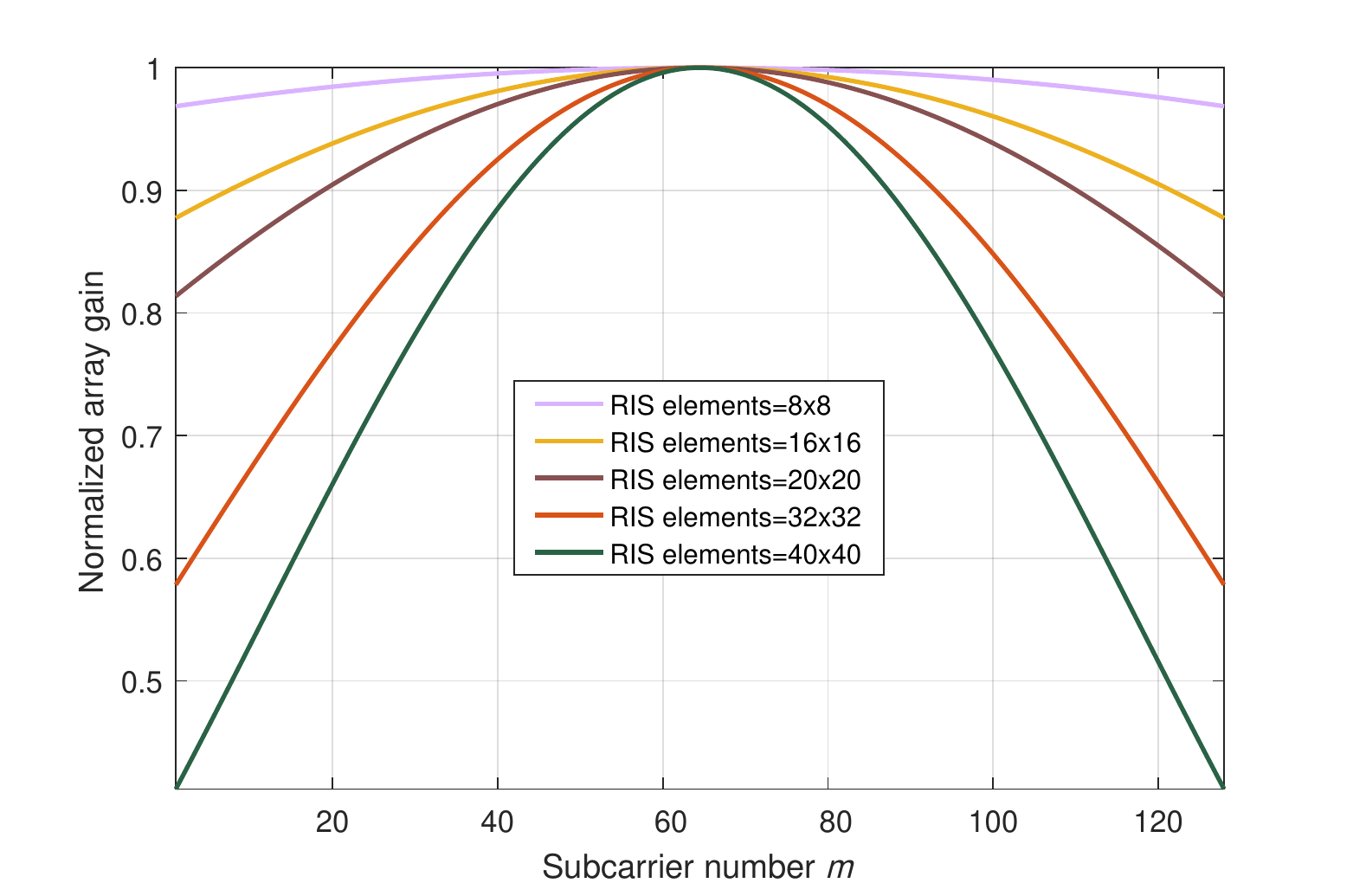}
	
	\caption{Normalized array gain under different RIS sizes.}
\end{figure}
\begin{figure}[htbp]  
\centering
    \label{RISrectangular} 
    \includegraphics[width=9cm,height=6cm]{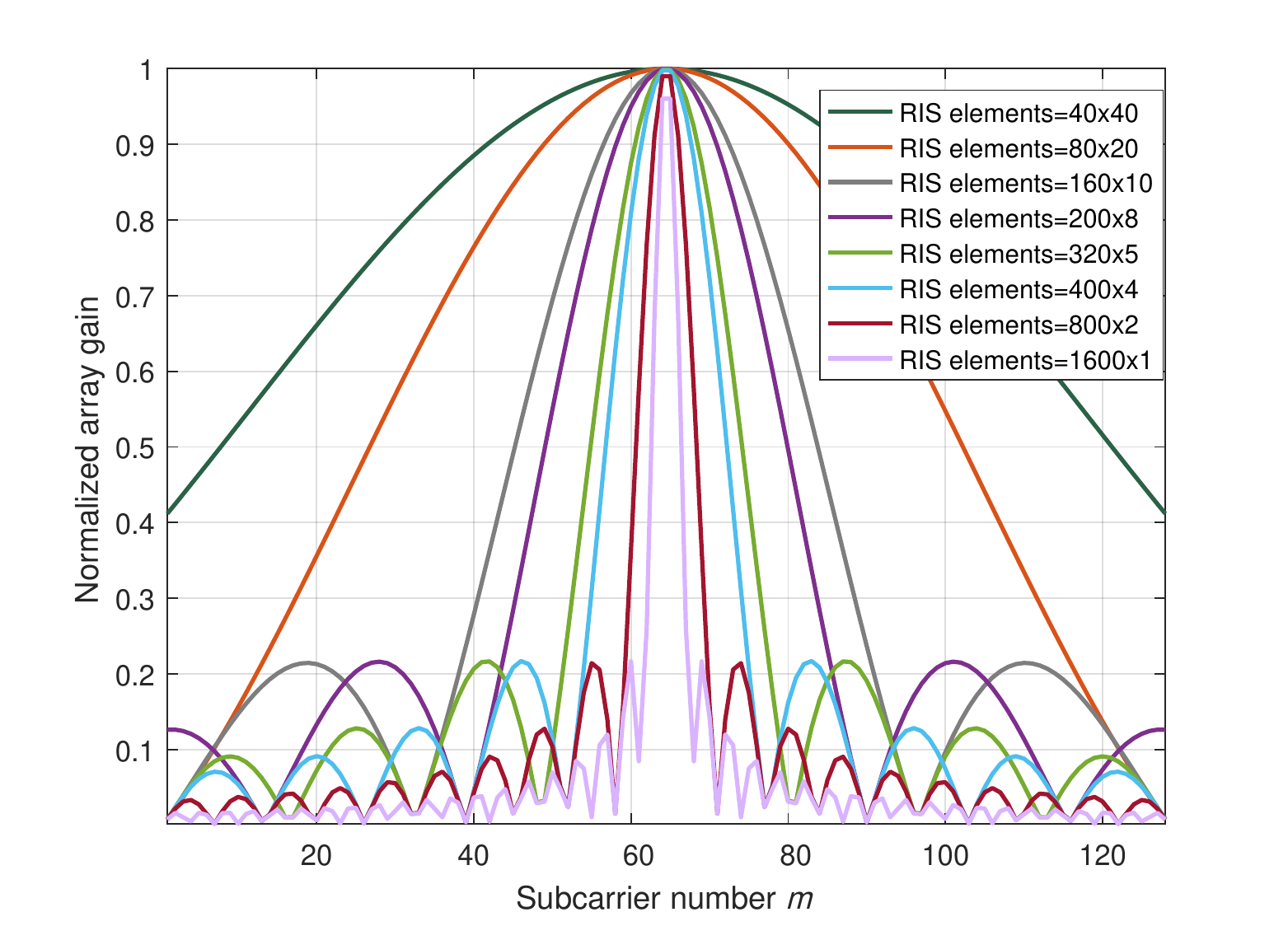}
	
	\caption{Normalized array gain under different RIS shapes.}
\end{figure}

In addition, it can be seen from (6) and (7) that more elements at the RIS lead to a larger signal delay, which further amplifies the beam  splits effects.  Generally, the RIS is a planar structure, and the horizon and elevation directions can both form the beam split, aggravating the effects. Thus, the centralized RIS can be replaced with small-size distributed RISs to alleviate the beam split. To further illustrate the effectiveness of distributed RIS, we analyze the condition that the centralized RIS is equally divided into $S$ distributed RISs.
The above deployment strategies may lead to different channel conditions, which results that the obtained normalized array gains are not based on the same condition. Therefore, we approximate that RISs are co-located, and the array gain for the distributed RIS deployment can be expressed as
\begin{eqnarray}\label{ZM}
\begin{aligned}
&\Gamma_{d}\left(f, u_{0}, v_{0}, \bar\phi,M_{s,x}, M_{s,y}\right)\\
&=\sum_{s=1}^{S}\left|\sum_{m_{s,x}=1}^{M_{s,x}} e^{j \pi\left\{\left(m_{s,x}-1\right) \hat{u}\right\}}
\sum_{m_{s,y}=1}^{M_{s,y}} e^{j \pi\left\{\left(m_{s,y}-1\right)\hat{v}\right\}}\right|\\
&=\sum_{s=1}^{S}\left|\frac{\sin \left(\frac{\pi M_{s,x}}{2} \hat{u}\right)}{\sin \left(\frac{\pi}{2} \hat{u}\right)} \frac{\sin \left(\frac{\pi M_{s,y}}{2} \hat{v}\right)}{\sin \left(\frac{\pi}{2} \hat{v}\right)}\right|.
\end{aligned}
\end{eqnarray}
Fig.~5 plots the normalized array gain under different RIS deployments, where we set $f_{c}=100$ GHz,~$B=10$ GHz, $M=128, (u_{0},v_{0})=(0.5,0.5)$. Scheme 1 is a $16\times 16$ centralized RIS deployment. Schemes 2 and 3 all includes 4 distributed RISs, their difference is that each one in Scheme 2 is a $8\times 8$ square RIS, while each one in Scheme 3 is a $16\times 4$ rectangular RIS. One can observe that the beam split effect under distributed RISs deployment is smaller than that under centralized RIS deployment. Additionally, we can find that the array gain under Scheme 2 is higher than that under Scheme 3. The results are consistent with Fig.~4.
\begin{figure}[htbp]
\centering
    \label{RIS2} 
    \includegraphics[width=9cm,height=6cm]{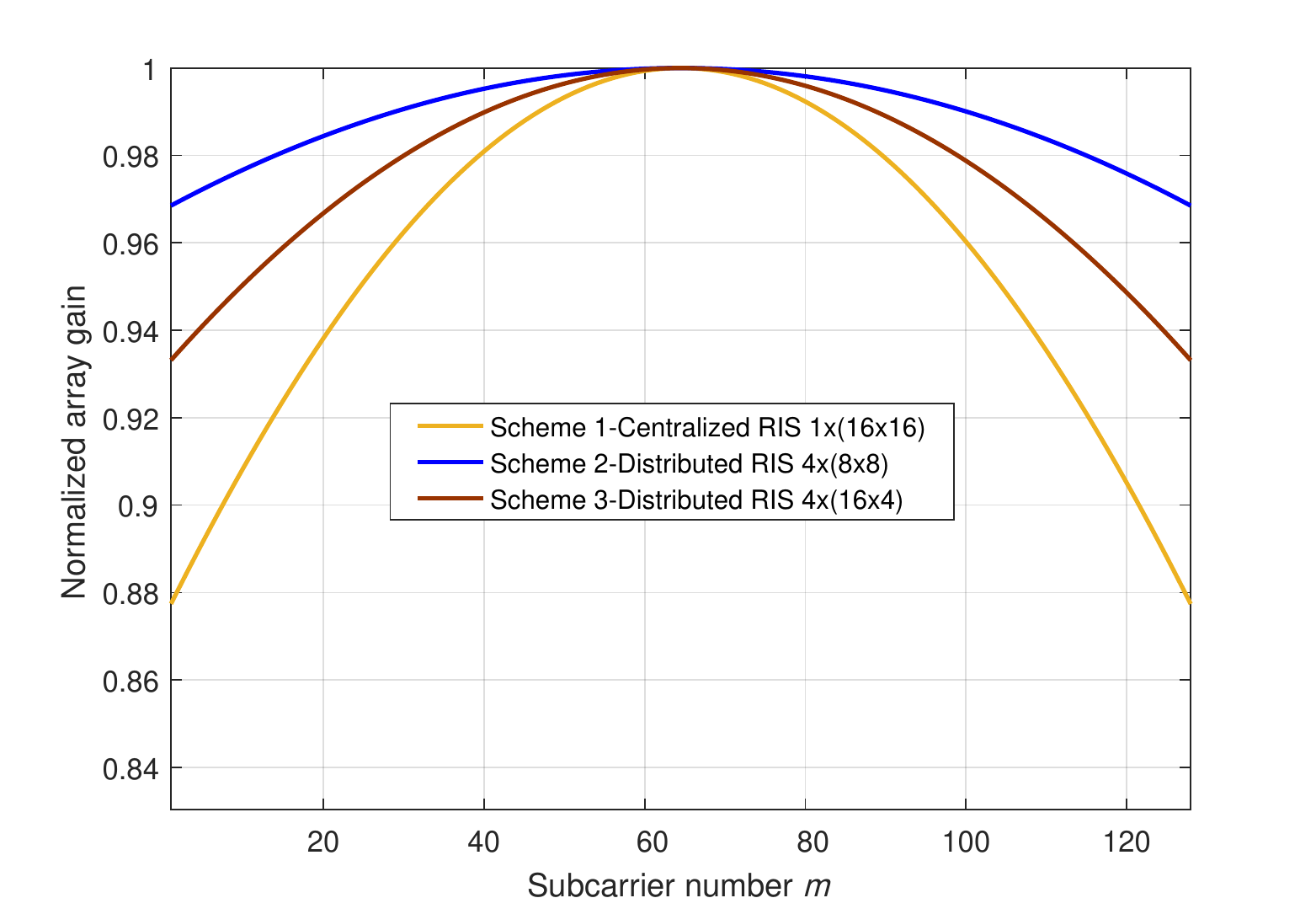}
	
	\caption{Normalized array gain under different RIS deployments.}
\end{figure}

According to the above analysis, we can find that the beam split effect at the RIS is related with the RIS sizes, shapes and deployments. The above conclusions can guide us how to solve or reduce  the beam split effect at the RIS side, and next we will investigate the beamforming design problem for the wideband THz RIS communications.

\section{System Model and Problem Formulation}
In this section, we investigate the beamforming design problem for the wideband  THz RIS communications. For overcoming the beam split effect, we adopt the FC-TD-PS-HB structure at the BS and distributed square RISs deployment scheme. For such scheme, we can apply the central control unit to manage the resource scheduling and allocation in the real system, and we can also obtain the channel state information based on the PARAFAC-based scheme\cite{ref102}.  On this basis, we study the beamforming design problem.
\subsection{System Model}
We assume that there are $R$ distributed RISs as shown in Fig. 1(b), and thus the equivalent channel $\mathbf{h}_{m, k}$ between the BS and the $k$-th user on the $m$-th subcarrier is expressed as
\begin{eqnarray}
  \mathbf{h}_{m, k}=\sum_{r=1}^{R} \mathbf{f}_{r, m, k} \mathbf{\Phi}_{r} \mathbf{G}_{r, m},
\end{eqnarray}
where $\mathbf{f}_{r, m, k} \in \mathbb{C}^{1 \times N_{\rm{RIS}}}$ denotes the channel between the $r$-th RIS and the $k$-th user on the $m$-th subcarrier, $\mathbf{G}_{r, m} \in \mathbb{C}^{N_{\rm{RIS}} \times N_{\rm{TX}}}$ represents the channel from the BS to the $r$-th RIS on the $m$-th subcarrier. $\mathbf{\Phi}_{r}=\operatorname{diag}\left(\varphi_{r,1,1}, \cdots, \varphi_{r,m_{x},m_{y}}, \cdots, \varphi_{r,M_{x},M_{y}}\right), r \in \mathcal{R}$, $m_{x} \in \mathcal{M}_{x}, m_{y} \in \mathcal{M}_{y}$ is the diagonal reflection coefficients matrix of the $r$-th RIS with $\varphi_{r,m_{x},m_{y}}= \varepsilon_{r,m_{x},m_{y}} e^{j \phi_{r,m_{x},m_{y}}}$. We define $\mathcal{R}=\{1, \cdots, R\}$ as the index set of RISs, and assume that all RISs own the same size. The channel matrix $\mathbf{G}_{r, m}$ from the BS to the $r$-th RIS and channel vector $\mathbf{f}_{r, m, k}$ from the $r$-th RIS to the $k$-th user on the $m$-th subcarrier can be expressed as
\begin{eqnarray}
\mathbf{G}_{r, m}=\sum_{l_{1}=1}^{L_{1}} \alpha_{r, l_{1}} e^{-j 2 \pi \tau_{l_{1}} f_{m}} \mathbf{b}\left(u_{l_{1}}^{r}, v_{l_{1}}^{r}\right) \mathbf{a}\left(\theta_{l_{1}}^{r}\right)^{ H},
\end{eqnarray}
and
\begin{eqnarray}
\mathbf{f}_{r, m, k}=\sum_{l_{2}=1}^{L_{2}} \alpha_{r, l_{2}} e^{-j 2 \pi \tau_{l_{2}} f_{m}} \mathbf{b}\left(u_{l_{2}}^{r, k}, v_{l_{2}}^{r, k}\right)^{T},
\end{eqnarray}
respectively.

As illustrated in Fig. 6, each RF chain is connected to $K_{\rm{T}}$ TD elements and each TD element is connected to $P = N_{\rm{TX}}/ K_{\rm{T}}$ frequency-independent PSs. Generally, the number of RF chains is larger than that of RISs, namely $N_{RF}\geq R$. For convenience, in this paper, we assume $N_{RF}=R$, and thus each RIS can be served by the analog beamforming generated by the unique RF chain~\cite{ref9}. The received signal of the $k$-th user on the $m$-th subcarrier can be written as
\begin{eqnarray}\label{ZP}
y_{m, k}=
\mathbf{h}_{m, k} \mathbf{F}_{\rm{A}} \mathbf{d}_{m, k} {s}_{m, k}+\sum_{j=1, j \neq k}^{K} \mathbf{h}_{m,k} \mathbf{F}_{\rm{A}} \mathbf{d}_{m, j} s_{m, j}+n_{m, k},
\end{eqnarray}
where $\mathbf{F}_{\rm{A}}=\mathbf{F} \mathbf{F}_{\rm{T}}$ and $\mathbf{F}\in \mathbb{C}^{N_{\rm{TX}} \times K_{\rm{T}}N_{\rm{RF}}}=[\mathbf{F}_{1}, \cdots, \mathbf{F}_{n}, \cdots,\mathbf{F}_{N_{\rm{RF}}}]$  is analog beamforming matrix,  $\mathbf{F}_{n}\in \mathbb{C}^{N_{\rm{TX}} \times K_{\rm{T}}}=\rm diag([\mathbf{c}_{n,1},\mathbf{c}_{n,2},\cdots,\mathbf{c}_{n,K_{\rm{T}}}])$ denotes the analog beamforming matrix generated by the PSs connecting to the $n$-th RF chain via TDs. $\mathbf{F}_{\rm{T}}\in \mathbb{C}^{ K_{\rm{T}}N_{\rm{RF}}\times N_{\rm{RF}}}=\operatorname{diag}\left(\left[e^{-j 2 \pi f_{m} \mathbf{t}_{1}}, e^{-j 2 \pi f_{m} \mathbf{t}_{2}},\cdots, e^{-j 2 \pi f_{m} \mathbf{t}_{N_{\rm{RF}}}}\right]\right)$ is the time delay matrix, where $\mathbf{t}_{n} \in \mathcal{C}^{K_{\rm{T}} \times 1}=\left[t_{n, 1}, t_{n, 2}, \cdots, t_{n, K_{\rm{T}}}\right]^{\rm T}$ is time delay vector realized by $K_{\rm{T}}$ TD elements connecting to the $n$-th RF chain. In addition,
$\mathbf{d}_{m, k}\in \mathbb{C}^{N_{\rm{RF}} \times 1}$ denotes digital beamforming vector. $n_{m,k} \sim \mathcal{C} \mathcal{N}\left(0, \sigma_{m,k}^{2}\right)$  is the additive zero average white Gaussian noise (AWGN) with variance of $\sigma_{m,k}^{2}$ at the $k$-th user on the $m$-th subcarrier, and $s_{m, k}$ denotes the transmit symbol to the $k$-th user on the $m$-th subcarrier with $E\left[\left|s_{m,k}\right|^{2}\right]=1$.

Then, the SINR of the $k$-th user on the $m$-th subcarrier can be calculated as
\begin{eqnarray}
\gamma_{m,k}=\frac{\left|\mathbf{h}_{m, k} \mathbf{F}_{\rm{A}} \mathbf{d}_{m, k}\right|^{2}}{\sum_{j=1, j \neq k}^{K}\left|\mathbf{h}_{m, k} \mathbf{F}_{\rm{A}} \mathbf{d}_{m, j}\right|^{2}+\sigma_{m, k}^{2}},
\end{eqnarray}
and the achievable rate is
\begin{eqnarray}
R_{\rm{sum}}=\sum_{k=1}^{K} \sum_{m=1}^{M} \log _{2}\left(1+\gamma_{m, k}\right).
\end{eqnarray}

\begin{figure}[t]
\centering
    \label{TD-PS2} 
    \includegraphics[width=8.5cm,height=6.5cm]{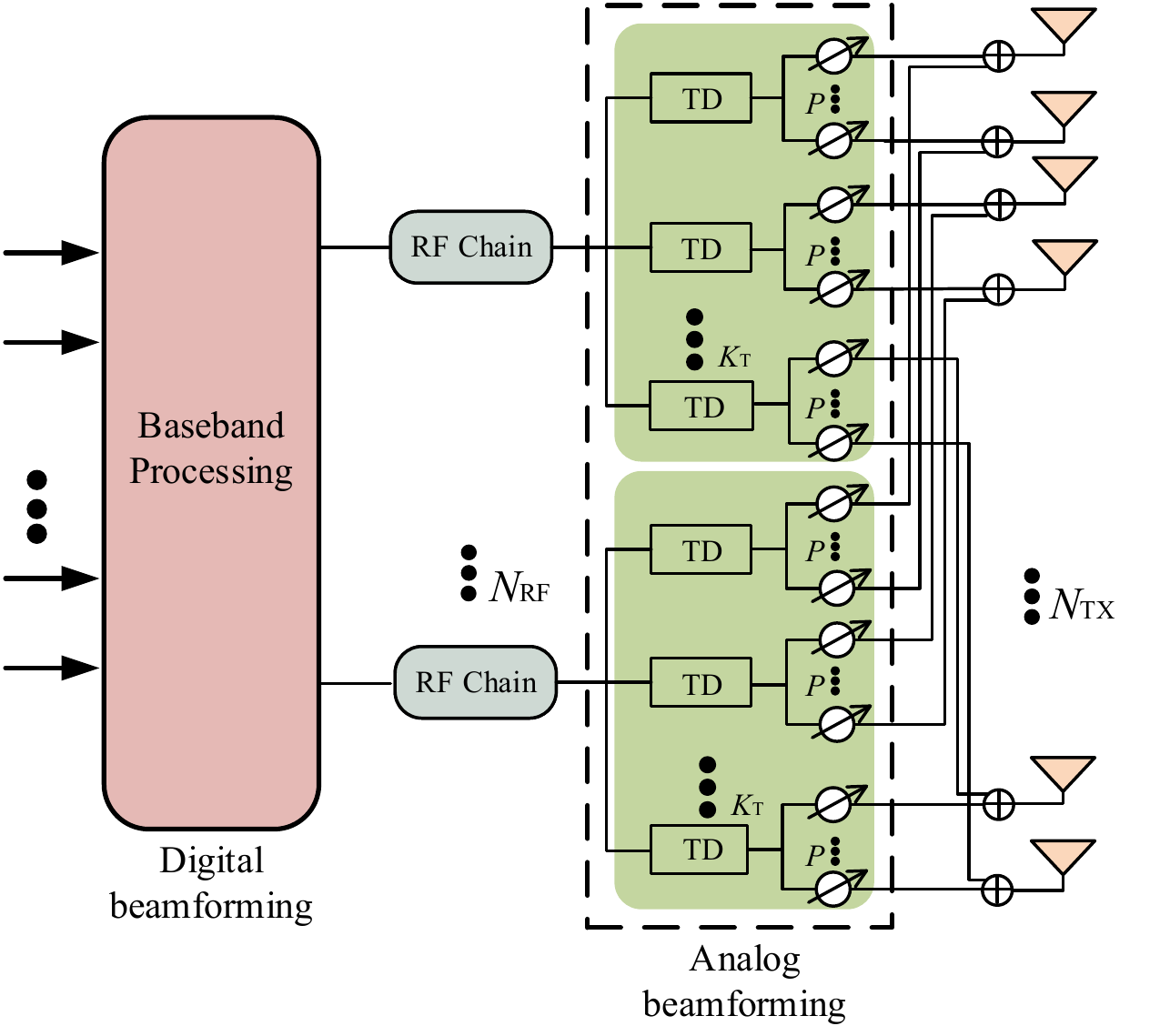}
	
	\caption{FC-TD-PS-HB architecture.}
\end{figure}
\subsection{Problem Formulation}
We assume that RISs can be controlled independently, and the reflection coefficients satisfy
\begin{eqnarray}
\mathcal{S} \triangleq\left\{\varphi_{r,m_{x},m_{y}}| | \varphi_{r,m_{x},m_{y}} \mid \leq 1\right\}, r \in \mathcal{R}, m_{x} \in \mathcal{M}_{x}, m_{y} \in \mathcal{M}_{y}.
\end{eqnarray}
Consequently, the optimization problem of maximizing the achievable sum rate can be formulated as
\begin{subequations}\label{OptA}
\begin{align}
\mathrm{P1}:&\max _{\boldsymbol{\Phi}, \mathbf{F}_{\rm{A}}, \mathbf{d}_{m, k}} R_{\rm{sum}}\label{OptA0}\\
{\rm{s.t.}}\;\;&\sum_{\mathrm{k}=1}^{K} \sum_{m=1}^{M}\left\|\mathbf{F}_{\rm{A}} \mathbf{d}_{m, k}\right\|^{2} \leq P_{\text {max }} \label{OptA1},\\
&\left|\varphi_{r, m_{x},m_{y}}\right|\leq 1, \forall r \in \mathcal{R}, m_{x} \in \mathcal{M}_{x}, m_{y} \in \mathcal{M}_{y}\label{OptA2},\\
&\left|\mathbf{c}_{n, k_{t}}\right|=\frac{1}{\sqrt{N_{\rm{T X}}}}, n=1,2, \ldots N_{\rm{RF}}, k_{t}=1,2, \ldots, K_{\rm{T}},
\end{align}
\end{subequations}
where $P_{\text {max }}$ is the maximum available transmit power and $\boldsymbol{\Phi}=\operatorname{diag}\left(\boldsymbol{\Phi}_{1}, \ldots, \boldsymbol{\Phi}_{\rm{R}}\right)$. P1 aims to jointly optimize the reflection coefficients matrix $\mathbf{\Phi}$, frequency-dependent analog beamforming matrix $\mathbf{F}_{\rm{A}}$ and digital beamforming vector $\mathbf{d}_{m, k}$ for maximizing the achievable sum rate. (32b) is the total transmit power constraint, (32c) is the constraint for each reflection coefficients, and (32d) is the amplitude constraint of analog beamforming. Due to the non-convex objective function (32a) and amplitude constraint (32d), the joint optimization problem is difficult to be solved directly. In the next section, we propose an effective algorithm to address it.
\section{Problem Solution}
To solve $\mathrm{P1}$, we first design the analog beamforming and time delays based on the different RISs' physical directions, and then propose an alternatively iterative optimization algorithm to obtain the digital beamforming and reflection coefficients.
\subsection{Optimization of $\mathbf{F}_{\rm{A}}$ ($\mathbf{F}$ and $ \mathbf{F}_{\rm{T}}$)}
We first design the frequency-dependent analog beamforming matrix $\mathbf{F}_{\rm{A}}=\mathbf{F} \mathbf{F}_{\rm{T}}$, which can be realized by PSs and TDs. Let the $n$-th column beamforming vector of $\mathbf{F}_{\rm{A}}$ to serve the $r$-th RIS on the $m$-th subcarrier, we define it as $\boldsymbol{\mathcal{F}}_{n,m}=\mathbf{F}_{n}e^{-j 2 \pi f_{m} \mathbf{t}_{n}}$.
According to~\cite{ref13},~\cite{ref29}, the PSs can be used to generate beams aligned with RISs' physical directions by adjusting their phase shifts, and the TDs can be used to rotate beams at different subcarriers to the RISs' physical directions by adjusting their time delays. Specifically, the array steering vector $\mathbf{a}\left(\eta_{l_{1},c}^{r}\right)$ is a constant-magnitude phase-only one which can be applied as columns of $\mathbf{F}$, where $\mathbf{F}=[\mathbf{F}_{1}, \cdots, \mathbf{F}_{n}, \cdots,\mathbf{F}_{N_{\rm{RF}}}]$ and $\mathbf{F}_{n}$ is given~by
\begin{eqnarray}\label{ZN}
\mathbf{F}_{n}=\operatorname{diag}\left(\left[\mathbf{a}_{1 \rightarrow P}\left(\eta_{l_{1},c}^{r}\right), \ldots, \mathbf{a}_{\left(K_{\rm T}-1\right) P \rightarrow K_{\rm T} P}\left(\eta_{l_{1},c}^{r}\right)\right]\right).
\end{eqnarray}
However, for most subcarriers, the beams generated by PSs cannot be aligned with the $r$-th RIS's spatial direction $\eta_{l_{1},c}^{r}$, where $\eta_{l_{1},c}^{r}$ is the spatial direction of the $l_{1}$-th path departing from the BS to the $r$-th RIS at the central frequency $f_{c}$. Furthermore, the beam on the $m$-th subcarrier will be aligned with the frequency-dependent spatial direction $\eta_{l_{1},m}^{r}$, which is
\begin{eqnarray}
\eta_{l_{1},m}^{r}=(f_{c}/f_{m}) \eta_{l_{1},c}^{r}.
\end{eqnarray}
To alleviate the beam split effect of the $n$-th beam transmitted to the $r$-th RIS, the TDs network is deployed to rotate the spatial direction from $\eta_{l_{1},m}^{r}$ to $\eta_{l_{1},c}^{r}$, which is realized by a extra phase shift $\zeta_{n,m}$, where $\zeta_{n,m}=(f_{m}/f_{c}-1)P\eta_{l_{1},c}^{r}$.
Consequently, for the $n$-th beam generated for the $r$-th RIS, the time delays should~be
\begin{eqnarray}\label{ZO}
e^{-j 2 \pi f_{m} \mathbf{t}_{n}}=\left[1, e^{j \pi \zeta_{n, m}}, e^{j \pi 2 \zeta_{n, m}}, \cdots, e^{j \pi(K_{\rm{T}}-1) \zeta_{n, m}}\right]^{T}.
\end{eqnarray}
And the time delays vector $\mathbf{t}_{n}$ can be expressed as
\begin{eqnarray}
\mathbf{t}_{n} \in \mathbb{C}^{K_{\rm{T}} \times 1}=\left[0, z_{n} T_{c}, \ldots, z_{n} T_{c}\left(K_{\rm{T}}-1\right)\right]^{T},
\end{eqnarray}
where $z_{n}=-\frac{P \sin \eta_{l_{1},c}^{r}}{2}$ represents the number of periods and $T_{c}$ is the period of the central~frequency.

Based on (\ref{ZN}) and (\ref{ZO}), the frequency-dependent beamforming vector $\boldsymbol{\mathcal{F}}_{n,m}=\mathbf{F}_{n}e^{-j 2 \pi f_{m} \mathbf{t}_{n}}$ can generate beam aligned with the $r$-th RIS's physical direction at all subcarriers. We can observe that $z_{n}$ only depends on $P$ and spatial direction $\eta_{l_{1},c}^{r}$ at the central frequency $f_{c}$. This denotes that an identical time delay can make up for the beam split over the whole bandwidth. Numerical results further demonstrate the effectiveness of the proposed scheme in Sec. V.
\subsection{Optimization of $\mathbf{d}_{m, k}$ with Fixed $\mathbf{\Phi}$}
After obtaining the analog beamforming matrix $\mathbf{F}_{\rm{A}}$, we optimize the digital beamforming vector $\mathbf{d}_{m, k}$ with fixed $\mathbf{\Phi}$. The equivalent channel vectors for the $k$-th user on the $m$-th subcarrier can be written as $\hat{\mathbf{h}}_{m, k}=\mathbf{h}_{m, k}\mathbf{F}_{\rm{A}}$. Besides, based on the extension of the Sherman-Morrison-Woodbury formula~\cite{ref15}
\begin{eqnarray}
(\mathbf{V}+\mathbf{X Y Z})^{-1}=\mathbf{V}^{-1}-\mathbf{V}^{-1} \mathbf{X}\left(\mathbf{I}+\mathbf{Y Z V}^{-1} \mathbf{X}\right)^{-1} \mathbf{Y Z V}^{-1},
\end{eqnarray}
we have
\begin{eqnarray}
\left(1+\gamma_{m, k}\right)^{-1}=1-\frac{\left|\hat{\mathbf{h}}_{m, k} \mathbf{d}_{m, k}\right|^{2}}{\sum_{j=1}^{K}\left|\hat{\mathbf{h}}_{m, k} \mathbf{d}_{m, j}\right|^{2}+\sigma_{m, k}^{2}}.
\end{eqnarray}
If MMSE detection technique~\cite{ref14} is used to obtain the original signal $s_{m,k}$ from $y_{m,k}$, the detection problem can be formulated~as
\begin{eqnarray}
u_{m, k}^{\rm opt}=\rm arg \min _{u_{m, k}} \epsilon_{m, k},
\end{eqnarray}
where
\begin{eqnarray}\label{ZQ}
\epsilon_{m, k}=\mathbb{E}\left[\left\|u_{m, k} y_{m, k}-s_{m, k}\right\|_{2}^{2}\right],
\end{eqnarray}
is the MSE, and $u_{m, k}$ is the channel equalization coefficient. Substituting (\ref{ZP}) into (\ref{ZQ}), we have
\begin{eqnarray}\label{ZR}
\begin{split}
\epsilon_{m, k}&=\sum_{j=1}^{K}\left|u_{m, k} \hat{\mathbf{h}}_{m, k} \mathbf{d}_{m, j}\right|^{2}-2 \operatorname{Re}\left\{u_{m, k} \hat{\mathbf{h}}_{m, k} \mathbf{d}_{m, k}\right\}+\left|u_{m, k}\right|^{2} \sigma_{m, k}^{2}+1.
\end{split}
\end{eqnarray}
Then, taking the partial derivatives to (\ref{ZR}) with respect to $u_{m, k}$ and setting the result to zero, the optimal equalization coefficient $u_{m, k}^{\rm opt}$ can be calculated as
\begin{eqnarray}\label{ZS}
u_{m, k}^{\rm opt}=\frac{\hat{\mathbf{h}}_{m, k} \mathbf{d}_{m, k}}{\sum_{j=1}^{K}\left|\hat{\mathbf{h}}_{m, k} \mathbf{d}_{m, j}\right|^{2}+\sigma_{m, k}^{2}}.
\end{eqnarray}
Substituting (\ref{ZS}) into (\ref{ZR}), the MMSE can be obtained as
\begin{eqnarray}
\epsilon_{m, k}^{\rm opt}=1-\frac{\left|\hat{\mathbf{h}}_{m, k} \mathbf{d}_{m, k}\right|^{2}}{\sum_{j=1}^{K}\left|\hat{\mathbf{h}}_{m, k} \mathbf{d}_{m, j}\right|^{2}+\sigma_{m, k}^{2}},
\end{eqnarray}
which is equal to $\left(1+\gamma_{m, k}\right)^{-1}$. Consequently, we have
\begin{eqnarray}
\left(1+\gamma_{m, k}\right)^{-1}=\min _{u_{m, k}} \epsilon_{m, k}.
\end{eqnarray}
Then, the achievable rate of the $k$-th user on the $m$-th subcarrier can be expressed as~\cite{add1}
\begin{eqnarray}\label{ZT}
\log _{2}\left(1+\gamma_{m, k}\right)=\max _{u_{m, k}}\left(-\log _{2} \epsilon_{m, k}\right).
\end{eqnarray}
Next, we give the following Proposition 1

$\textit{Proposition 1}$: Let $f(\mathcal{P})=-\frac{\mathcal{P} \mathcal{Q}}{\ln 2}+\log _{2} \mathcal{P}+\frac{1}{\ln 2}$ and $\mathcal{P}$ be a positive real number, we have $\max _{\mathcal{P}>0} f(\mathcal{P})=-\log _{2} \mathcal{Q}$, where the optimal $\mathcal{P}$ is $\mathcal{P}^{\rm opt}=\frac{1}{\mathcal{Q}}$.

After removing the log function of (\ref{ZT}), based on~\cite{ref16},~\cite{ref17} and Proposition 1, we have
\begin{eqnarray}
\log _{2}\left(1+\gamma_{m, k}\right)=\max _{u_{m, k}} \max _{\tau_{m, k}>0}\left(-\frac{\tau_{m, k} \epsilon_{m, k}}{\ln 2}+\log _{2} \tau_{m, k}+\frac{1}{\ln 2}\right),
\end{eqnarray}
where the optimal $\tau_{m, k}$ is $\tau_{m, k}^{\rm opt}=\frac{1}{\epsilon_{m, k}}$.

Next, $\mathrm{P1}$ can be transformed as the MSE minimization problem, namely
\begin{subequations}\label{OptA}
\begin{align}
\mathrm{P2}:&\max _{\mathbf{d}_{m, k}} \sum_{k=1}^{K} \sum_{m=1}^{M} \max _{u_{m, k}} \max _{\tau_{m, k}>0}\left(-\frac{\tau_{m, k} \epsilon_{m, k}}{\ln 2}+\log _{2} \tau_{m, k}+\frac{1}{\ln 2}\right)\label{OptA0}\\
{\rm{s.t.}}\;\;&\sum_{k=1}^{K} \sum_{m=1}^{M}\left\|\mathbf{F}_{\rm{A}} \mathbf{d}_{m, k}\right\|^{2} \leq P_{\text {max }} \label{OptA1}.
\end{align}
\end{subequations}
To solve $\mathrm{P2}$, an iterative optimization algorithm is proposed. Based on the obtained $\mathbf{d}_{m, k}^{(t-1)}$ at the $(t-1)$-th iteration, $u_{m, k}^{(t)}$ at the $t$-th iteration can be expressed as
\begin{eqnarray}
u_{m, k}^{(t)}=\frac{\hat{\mathbf{h}}_{m, k} \mathbf{d}_{m, k}^{(t-1)}}{\sum_{j=1}^{K}\left|\hat{\mathbf{h}}_{m, k} \mathbf{d}_{m, j}^{(t-1)}\right|^{2}+\sigma_{m, k}^{2}}.
\end{eqnarray}
And the optimal $\tau_{m, k}^{(t)}$ at the $t$-th iteration can be calculated as $\tau_{m, k}^{(t)}=\frac{1}{\epsilon_{m, k}^{\rm {opt}(t)}}$
, where
\begin{eqnarray}
\epsilon_{m, k}^{\rm {opt}(t)}=1-\frac{\left|\hat{\mathbf{h}}_{m, k} \mathbf{d}_{m, k}^{(t-1)}\right|^{2}}{\sum_{j=1}^{K}\left|\hat{\mathbf{h}}_{m, k} \mathbf{d}_{m, j}^{(t-1)}\right|^{2}+\sigma_{m, k}^{2}}.
\end{eqnarray}
Then, $\mathrm{P2}$ is transformed as
\begin{subequations}\label{OptA}
\begin{align}
\;\;\mathrm{P3}:&\min _{\mathbf{d}_{m, k}^{(t)}} \sum_{k=1}^{K} \sum_{m=1}^{M} \left(\frac{\tau_{m, k}^{(t)} \epsilon_{m, k}^{(t)}}{\ln 2}-\log _{2} \tau_{m, k}^{(t)}-\frac{1}{\ln 2}\right)\label{OptA0}\\
{\rm{s.t.}}\;\;&\sum_{k=1}^{K} \sum_{m=1}^{M}\left\|\mathbf{F}_{\rm{A}} \mathbf{d}_{m, k}^{(t)}\right\|^{2} \leq P_{\text {max }} \label{OptA1},
\end{align}
\end{subequations}
where
\begin{eqnarray}
\begin{split}
\epsilon_{m, k}^{(t)}&=\sum_{j=1}^{K}\left|u_{m, k}^{(t)} \hat{\mathbf{h}}_{m, k} \mathbf{d}_{m, j}^{(t)}\right|^{2}-2 \operatorname{Re}\left\{u_{m, k}^{(t)} \hat{\mathbf{h}}_{m, k} \mathbf{d}_{m, k}^{(t)}\right\}+\left|u_{m, k}^{(t)}\right|^{2} \sigma_{m, k}^{2}+1.
\end{split}
\end{eqnarray}
It is obvious that $\mathrm{P3}$ is a standard convex optimization problem, which can be solved by numerical convex program solvers~\cite{ref18}. Particularly, since the obtained $\mathbf{d}_{m, k}^{(t)}$, $\tau_{m, k}^{(t)}$, $u_{m, k}^{(t)}$ are the optimal solutions of $\mathrm{P3}$ at the $t$-th iteration, iteratively updating these variables will increase or maintain the value of the objective function in $\mathrm{P3}$~\cite{ref14}. Consequently, the proposed alternatively iterative optimization scheme will converge to at least a local optimal solution.
\subsection{Optimization of $\mathbf{\Phi}$ with Fixed $\mathbf{d}_{m, k}$}
Based on the obtained analog beamforming matrix $\mathbf{F}_{A}$ and digital beamforming  vector $\mathbf{d}_{m, k}$, we define $\mathbf{w}_{m, k}=\mathbf{F}_{\rm{A}} \mathbf{d}_{m, k}$ as the equivalent beamforming. Next, we optimize the reflection coefficients matrix $\mathbf{\Phi}$. To solve the logarithms in the objective function of $\mathrm{P1}$, we apply the LDR method \cite{add2} and introduce an auxiliary variable $\boldsymbol{\rho}=\left[\mathbf{\rho}_{1,1}, \mathbf{\rho}_{1,2}, \ldots, \mathbf{\rho}_{1, K}, \mathbf{\rho}_{2,1}, \mathbf{\rho}_{2,2}, \ldots, \mathbf{\rho}_{\rm{M, K}}\right]^{T}$. Then, the objective function can be expressed as
\begin{eqnarray}
\begin{split}
f(\mathbf{\Phi}, \mathbf{W}, \boldsymbol{\rho})=\sum_{k=1}^{K} \sum_{m=1}^{M} \ln \left(1+\rho_{m, k}\right)-\sum_{k=1}^{K} \sum_{m=1}^{M} \rho_{m, k}+
\sum_{k=1}^{K} \sum_{m=1}^{M}\left(1+\rho_{m, k}\right) f_{m, k}(\mathbf{\Phi}, \mathbf{W}),
\end{split}
\end{eqnarray}
where $\mathbf{W}=\left[\mathbf{w}_{1,1}^{T}, \mathbf{w}_{1,2}^{T}, \ldots, \mathbf{w}_{1, K}^{T}, \mathbf{w}_{2,1}^{T}, \mathbf{w}_{2,2}^{T}, \ldots, \mathbf{w}_{\rm{M, K}}^{T}\right]^{T}$, and $f_{k, m}(\mathbf{\Phi}, \mathbf{W})$ is denoted as
\begin{eqnarray}
f_{m, k}(\mathbf{\Phi}, \mathbf{W})=\frac{\left|\mathbf{h}_{m, k} \mathbf{w}_{m, k}\right|^{2}}{\sum_{j=1}^{K}\left|\mathbf{h}_{m, k} \mathbf{w}_{m, j}\right|^{2}+\sigma_{m, k}^{2}}.
\end{eqnarray}
Consequently, $\mathrm{P1}$ can be transformed as
\begin{subequations}\label{OptA}
\begin{align}
\;\;\mathrm{P4}:&\max _{\mathbf{\Phi}, \boldsymbol{\rho}} f(\mathbf{\Phi}, \mathbf{W}, \boldsymbol{\rho})\label{OptA0}\\
{\rm{s.t.}}\;\;&\left|\varphi_{r, m_{x},m_{y}}\right|\leq 1,
\forall r \in \mathcal{R}, m_{x} \in \mathcal{M}_{x}, m_{y} \in \mathcal{M}_{y}\label{OptA1}.
\end{align}
\end{subequations}
Since it is difficult to optimize $\mathbf{\Phi}$ and $\boldsymbol{\rho}$ simultaneously, we propose an alternatively iterative optimization technique to solve it.

Firstly, for given $(\mathbf{\Phi}^{*},\mathbf{W}^{*})$, it is obvious that (52) is a concave function of $\boldsymbol{\rho}$, and thus $\boldsymbol{\rho}$ can be directly obtained by setting $\partial f / \partial \rho_{m, k}=0$ for $\forall k \in \left[1,2, \cdots, K\right]$, $\forall m \in \left[1,2, \cdots, M\right]$, i.e.,
\begin{eqnarray}\label{ZA}
\begin{split}
\rho_{m, k}^{\rm opt}=\gamma_{m, k}^{*}=\frac{\left|\mathbf{h}_{m, k} \mathbf{w}_{m, k}\right|^{2}}{\sum_{j=1, j \neq k}^{K}\left|\mathbf{h}_{m, k} \mathbf{w}_{m, j}\right|^{2}+\sigma_{m, k}^{2}}, \\
\forall k \in \left[1,2, \cdots, K\right], \forall m \in \left[1,2, \cdots, M\right].
\end{split}
\end{eqnarray}
After obtaining $(\boldsymbol{\rho}^{*}, \mathbf{W}^{*})$, $\mathrm{P4}$ can be transformed as
\begin{subequations}\label{OptA}
\begin{align}
\;\;\mathrm{P5}:&\max _{\mathbf{\Phi}}\sum_{k=1}^{K} \sum_{m=1}^{M}\left(1+\rho_{ m,k}^{*}\right) f_{m, k}(\mathbf{\Phi}, \mathbf{W}^{*})\label{OptA0}\\
{\rm{s.t.}}\;\;&\left|\varphi_{r, m_{x},m_{y}}\right|\leq 1,
\forall r \in \mathcal{R}, m_{x} \in \mathcal{M}_{x}, m_{y} \in \mathcal{M}_{y}\label{OptA1}.
\end{align}
\end{subequations}
To simplify the expression of (56a), we introduce a auxiliary function with respect to $\mathbf{\Phi}$, which is expressed as
\begin{eqnarray}
\mathbf{Q}_{k, m, j}(\boldsymbol{\Phi})=\mathbf{f}_{m, k} \boldsymbol{\Phi} \mathbf{G}_{m} \mathbf{w}_{m, j},
\end{eqnarray}
where $\boldsymbol{\Phi}=\operatorname{diag}\left(\boldsymbol{\Phi}_{1}, \ldots, \boldsymbol{\Phi}_{\rm{R}}\right)$, $\mathbf{G}_{m}=[\mathbf{G}_{1,m}^{\rm T}, \mathbf{G}_{2, m}^{\rm T}, \cdots, \mathbf{G}_{R,m}^{\rm T}]^{\rm T}$, $\mathbf{f}_{k, m}=[\mathbf{f}_{1,m,k}, \mathbf{f}_{2,m,k}, \cdots, \mathbf{f}_{R,m,k}]$.
Consequently, $\mathrm{P5}$ can be re-expressed as
\begin{subequations}\label{OptA}
\begin{align}
\;\;\mathrm{P6}:&\max _{\mathbf{\Phi}} \Upsilon_{1}=\sum_{k=1}^{K} \sum_{m=1}^{M} \Omega_{1}(\mathbf{\Phi})\label{OptA0}\\
{\rm{s.t.}}\;\;&\left|\varphi_{r, m_{x},m_{y}}\right|\leq 1,
\forall r \in \mathcal{R}, m_{x} \in \mathcal{M}_{x}, m_{y} \in \mathcal{M}_{y}\label{OptA1},
\end{align}
\end{subequations}
where
\begin{eqnarray}
\Omega_{1}(\mathbf{\Phi})=(1+\rho_{m, k}^{*}) \frac{\left|\mathbf{Q}_{k, m, k}(\boldsymbol{\Phi})\right|^{2}}{\sum_{j=1}^{K}\left|\mathbf{Q}_{k, m, j}(\boldsymbol{\Phi})\right|^{2}+\sigma_{m, k}^{2}}.
\end{eqnarray}
However, $\mathrm{P6}$ is still nontrivial to solve due to the multidimensional fractions in (58a). Fortunately,  $\mathrm{P6}$ satisfies the concave-convex conditions~\cite{ref19}, then we can exploit the multidimensional complex quadratic transform (MCQT) technique. Different from the common fractional programming (FP)-based method, MCQT extends the common scalar-form FP to matrix-form, which can be used to solve the non-convexity of the high-dimensional fractions. Thus, we introduce an auxiliary variable $\boldsymbol{\chi}=\left[\mathbf{\chi}_{1,1}, \mathbf{\chi}_{1,2}, \ldots, \mathbf{\chi}_{1, K}, \mathbf{\chi}_{2,1}, \mathbf{\chi}_{2,2}, \ldots, \mathbf{\chi}_{M, K}\right]^{T}$ and $\mathrm{P6}$ can be reformulated as
\begin{subequations}\label{OptA}
\begin{align}
\;\;\mathrm{P7}:&\max _{\mathbf{\Phi}, \boldsymbol{\chi}}\Upsilon_{2}=\sum_{k=1}^{K} \sum_{m=1}^{M} \Omega_{2}(\mathbf{\Phi}, \boldsymbol{\chi})\label{OptA0}\\
{\rm{s.t.}}\;\;&\left|\varphi_{r, m_{x},m_{y}}\right|\leq 1,
\forall r \in \mathcal{R}, m_{x} \in \mathcal{M}_{x}, m_{y} \in \mathcal{M}_{y}\label{OptA1},
\end{align}
\end{subequations}
where
\begin{eqnarray}
\begin{split}
\Omega_{2}(\mathbf{\Phi}, \boldsymbol{\chi})=2 \sqrt{1+\rho_{m, k}^{*}} \operatorname{Re}\left\{\mathbf{\chi}_{m, k}^{ H} \mathbf{Q}_{k, m, k}(\boldsymbol{\Phi})\right\}-\mathbf{\chi}_{m, k}^{ H}\left(\sum_{j=1}^{K}\left|\mathbf{Q}_{k, m, j}(\boldsymbol{\Phi})\right|^{2}+\sigma_{m, k}^{2}\right) \mathbf{\chi}_{m, k}.
\end{split}
\end{eqnarray}
Next, $\mathrm{P7}$ is divided into two subproblems and we optimize the reflection coefficients matrix $\mathbf{\Phi}$ and auxiliary variable $\boldsymbol{\chi}$ alternatively. The optimal $\mathbf{\chi}_{m, k}$ can be obtained by setting $\partial \Upsilon_{2} / \partial \mathbf{\chi}_{m, k}=0$ for $\forall k \in \left[1,2, \cdots, K\right]$, $\forall m \in \left[1,2, \cdots, M\right]$, under a given $\mathbf{\Phi}$, i.e.,
\begin{eqnarray}\label{ZB}
\mathbf{\chi}_{m, k}^{opt}=\sqrt{1+\rho_{m, k}^{*}} \frac{\mathbf{Q}_{k, m, k}(\boldsymbol{\Phi})}{\sum_{j=1}^{K}\left|\mathbf{Q}_{k, m, j}(\boldsymbol{\Phi})\right|^{2}+\sigma_{m, k}^{2}}.
\end{eqnarray}
After obtaining $\boldsymbol{\chi}$, we only need to optimize $\mathbf{\Phi}$, and thus $\Upsilon_{2}$ can be simplified as
\begin{eqnarray}
\begin{aligned}
\mathbf{\chi}_{m, k}^{ H} \mathbf{Q}_{k, m, j}(\boldsymbol{\Phi})&=
\mathbf{\chi}_{m, k}^{ H} \mathbf{f}_{m, k} \boldsymbol{\Phi} \mathbf{G}_{m} \mathbf{w}_{m, j}\\
&=\boldsymbol{\psi}^{H} \operatorname{daig}\left(\mathbf{\chi}_{m, k}^{ H} \mathbf{f}_{m, k}\right) \mathbf{G}_{m} \mathbf{w}_{m, j}\\
&=\boldsymbol{\psi}^{H} \mathbf{q}_{k, m, j},
\end{aligned}
\end{eqnarray}
where $\boldsymbol{\psi}=[\psi_{1},\cdots, \psi_{r}, \cdots, \psi_{\rm{R}}]^{\rm T}$, $\psi_{r}=[\varphi_{r,1,1}, \cdots, \varphi_{r,M_{x},M_{y}}]$, $r \in \left[1,2, \cdots, R\right]$ and we define $\mathbf{q}_{k, m, j}=\operatorname{daig}\left(\mathbf{\chi}_{m, k}^{H} \mathbf{f}_{m, k}\right) \mathbf{G}_{m} \mathbf{w}_{m, j}$. Substituting (63) into (61), which is equivalently transformed to a new function of $\boldsymbol{\psi}$, we have
\begin{eqnarray}
\begin{split}
\Omega_{2}(\boldsymbol{\psi})=2 \sqrt{1+\rho_{m, k}^{*}} \operatorname{Re}\left\{\boldsymbol{\psi}^{ H} \mathbf{q}_{k, m, j}\right\}
-\sum_{j=1}^{K}(\boldsymbol{\psi}^{ H} \mathbf{q}_{k, m, j})( \mathbf{q}_{k, m, j}^{ H}\boldsymbol{\psi}){\color{blue}-}\mathbf{\chi}_{m, k}^{ H}\sigma_{m, k}^{2} \mathbf{\chi}_{m, k}.
\end{split}
\end{eqnarray}
Substituting (64) into (60a), $\Upsilon_{2}$ is represented as
\begin{eqnarray}
\Upsilon_{2}=-\boldsymbol{\psi}^{ H} \Lambda \boldsymbol{\psi}+\operatorname{Re}\left\{2 \boldsymbol{\psi}^{ H} \upsilon\right\}-\varsigma,
\end{eqnarray}
where
\begin{eqnarray}
\Lambda=\sum_{k=1}^{K} \sum_{m=1}^{M} \sum_{j=1}^{K} \mathbf{q}_{k, m, j} \mathbf{q}_{k, m, j}^{ H},
\end{eqnarray}
\begin{eqnarray}
\upsilon=\sum_{k=1}^{K} \sum_{m=1}^{M} \sqrt{1+\rho_{m, k}^{*}} \mathbf{q}_{k, m, k},
\end{eqnarray}
\begin{eqnarray}
\varsigma=\sum_{k=1}^{K} \sum_{m=1}^{M} \mathbf{\chi}_{m, k}^{ H} \sigma_{m, k}^{2} \mathbf{\chi}_{m, k}.
\end{eqnarray}
Consequently, $\mathrm{P7}$ can be rewritten as
\begin{subequations}\label{ZC}
\begin{align}
\;\;\mathrm{P8}:&\min _{\boldsymbol{\psi}}\Upsilon_{3}=\boldsymbol{\psi}^{ H} \Lambda \boldsymbol{\psi}-\operatorname{Re}\left\{2 \boldsymbol{\psi}^{ H} \upsilon\right\}\label{OptA0}\\
{\rm{s.t.}}\;\;&\left|\varphi_{r, m_{x},m_{y}}\right|\leq 1,
\forall r \in \mathcal{R}, m_{x} \in \mathcal{M}_{x}, m_{y} \in \mathcal{M}_{y}\label{OptA1}.
\end{align}
\end{subequations}
It is obvious that $\mathbf{q}_{k, m, j} \mathbf{q}_{k, m, j}^{ H}$ is a positive-definite matrix for any $k$ and $m$, $\Lambda$ is a positive-definite matrix, $\Upsilon_{3}$ is a quadratic concave function of $\boldsymbol{\psi}$, and (69b) are convex constraints. Thus, $\mathrm{P8}$ is a standard QCQP problems and can be solved by ADMM~\cite{add4}. According to~\cite{ref19}, each step of the iteration, i.e., (\ref{ZA}), (\ref{ZB}), (\ref{ZC}), can be easily proved to be monotonous. Therefore, the algorithm to optimize the reflection coefficient of RIS has strict convergency. Moreover, simulation results in Section V also further verify the convergence of the proposed algorithm.

Finally, we summarize the proposed optimization scheme for solving $\mathrm{P1}$ in $\bf Algorithm \hspace*{0.02in} 1$. Specifically, the analog beamforming matrix $\mathbf{F}_{\rm{A}}$ is firstly designed according to (33) and (35). Next, based on the initially feasible digital beamforming vector $\mathbf{d}_{m, k}^{(0)}$ and reflection coefficients matrix $\mathbf{\Phi}^{(0)}$, the digital beamforming vector $\mathbf{d}_{m, k}^{(t)}$ at the $t$-th iteration can be obtained via the MMSE approach. Then, based on obtained $\mathbf{d}_{m, k}^{(t)}$, the reflection coefficients matrix $\mathbf{\Phi}^{(t)}$ at the $t$-th iteration is solved via LDR and MCQT methods. The above steps are repeated until convergence.
\begin{algorithm}
\caption{The Proposed Algorithm for Solving $\mathrm{P1}$}
{\bf Input:}
Channels $\mathbf{f}_{r, m, k}$, $\mathbf{G}_{r, m}$.\\
{\bf Initialization:}
Digital beamforming vector $\mathbf{d}_{m, k}^{(0)}$ and reflection coefficients matrix $\mathbf{\Phi}^{(0)}$.

Calculate the analog beamforming matrix $\mathbf{F}_{\rm{A}}$ according to (33) and (35).\\
{\bf while} no convergence {\bf do}\\
Obtain the digital beamforming vector $\mathbf{d}_{m, k}$ via solving P3;\\
Update the variable $\mathbf{\boldsymbol{\rho}}$ based on (55);\\
Update the variable $\mathbf{\boldsymbol{\chi}}$ based on (62);\\
Obtain the reflection coefficients matrix $\mathbf{\Phi}$ via solving P8;\\
{\bf end while}

{\bf Output:}
Analog beamforming matrix $\mathbf{F}_{\rm{A}}$, digital \hspace*{0.02in} beamforming vector $\mathbf{d}_{m, k}$, reflection coefficients matrix $\mathbf{\Phi}$.
\end{algorithm}
\subsection{Computational Complexity}
Now, we analyze the computational complexity for the proposed $\bf Algorithm \hspace*{0.02in} 1$. In fact, solving the digital beamforming vector $\mathbf{d}_{m, k}$, auxiliary variables $\mathbf{\boldsymbol{\rho}}$, $\mathbf{\boldsymbol{\chi}}$ and reflection coefficients matrix $\mathbf{\Phi}$ take up the dominant computation cost. Specifically, the dominant term in computational complexity of the MMSE approach to obtain the digital beamforming vector $\mathbf{d}_{m, k}$ is $\mathcal{O}\left(M N_{\rm{TX}}^{2}\right)$.
Computing the auxiliary variables $\mathbf{\boldsymbol{\rho}}$ and $\mathbf{\boldsymbol{\chi}}$ involves the complexity $\mathcal{O}\left(KM(KN_{\rm{TX}}+K+1)\right)$ and $\mathcal{O}\left(KM(K+1)\right)$, respectively.
Finally, the computational complexity is $\mathcal{O}\left(R^{3} N_{\rm{RIS}}^{3}\right)$ for solving the reflection coefficients matrix $\mathbf{\Phi}$, which is mainly caused by the matrix inversion operation. The overall computational complexity of the proposed $\bf Algorithm \hspace*{0.02in} 1$ is $\mathcal{O}\left(I_{\rm{o}} (R^{3} N_{\rm{RIS}}^{3}+M N_{\rm{TX}}^{2}+KM(KN_{\rm{TX}}+K+1)+KM(K+1))\right)$, where $I_{o}$ is the required number of iterations.
\section{Numerical Results}
In this section, simulation results are presented to evaluate the performance of the proposed scheme. To relieve the beam split effect, 4 distributed small-size RISs are deployed, and their locations are (0, 80 m, 6 m), (0, 80 m, 8 m), (0, 100 m, 6 m), (0, 100 m, 8 m), respectively, as shown in Fig.~7. Meanwhile, we also provide the comparison scheme based on one centralized large-size RIS deployment scheme. We assume that there are 4 users, and they are randomly distributed in a circle centered at (0, 85 m, 0) with radius of 1m. The default simulation parameters are listed in Table I.

\begin{table}[htb]
\begin{center}
\caption{System parameters}
\label{table:1}
\begin{tabular}{|c|c|c|}
\hline   \textbf{Parameters} & \textbf{Value} \\
\hline   Central frequency & $f_{c}=100$ GHz  \\
\hline   Bandwidth & $B=10$ GHz  \\
\hline   Number of subcarriers & $M=8$  \\
\hline   Number of TDs &  $K_{\rm{T}}=16$ \\
\hline   Number of RISs &  $R=4$ \\
\hline   Number of RIS elements &  $N_{\rm{RIS}}=64$ \\
\hline   Number of users &  $K=4$ \\
\hline   Number of RF chains & $N_{\rm{RF}}=4$  \\
\hline   Maximum transmit power & $P_{\rm{max}}=0$ dBm  \\
\hline   Noise power & $\sigma_{m,k}^{2} = -82$ dBm  \\
\hline   Number of paths & $L_{1}=L_{2}= 1$  \\
\hline
\end{tabular}
\end{center}
\end{table}

\begin{figure}[htbp]
\centering
    \label{location} 
    \includegraphics[width=9cm,height=6cm]{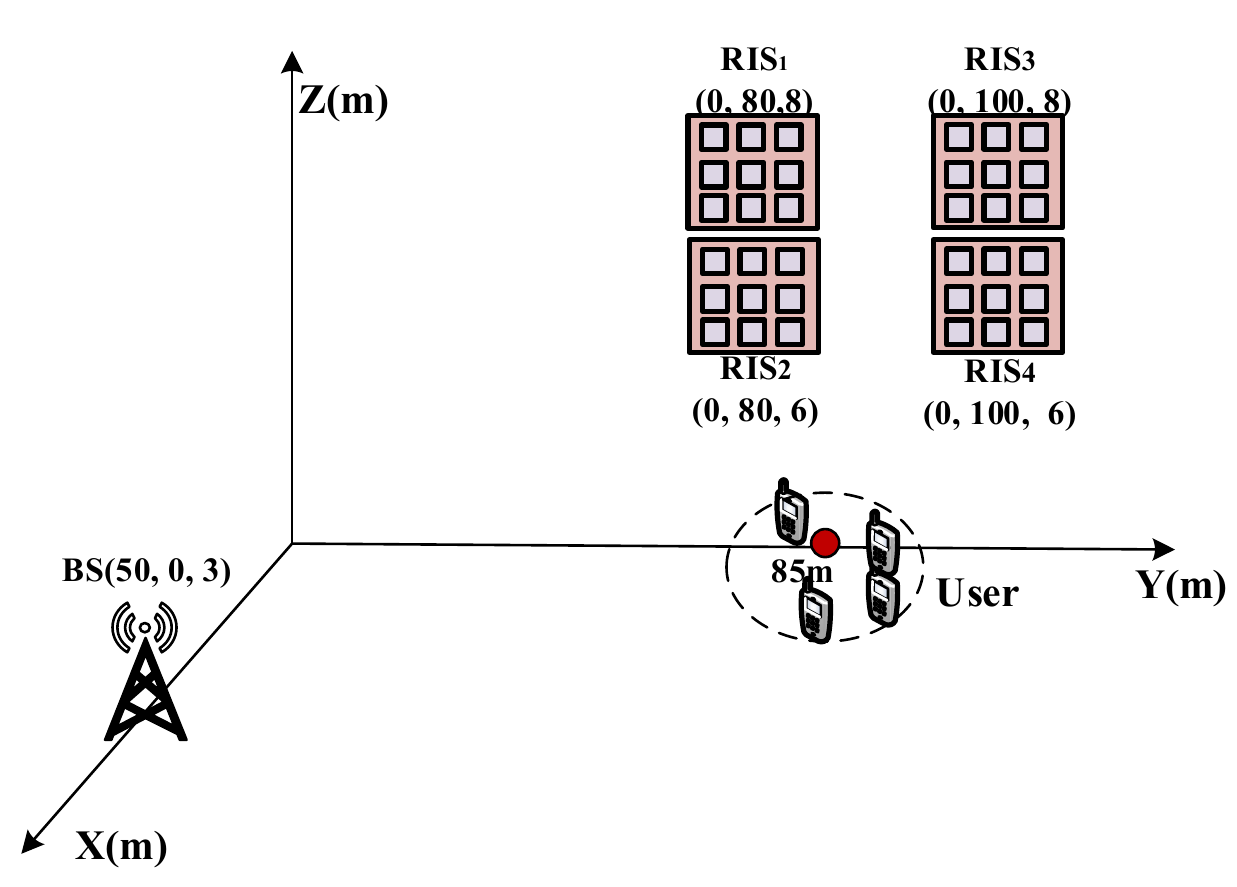}
	
	\caption{The location distribution for the considered system.}
\end{figure}

\begin{figure}[htbp]
\centering
    \label{fully-digital} 
    \includegraphics[width=9cm,height=6cm]{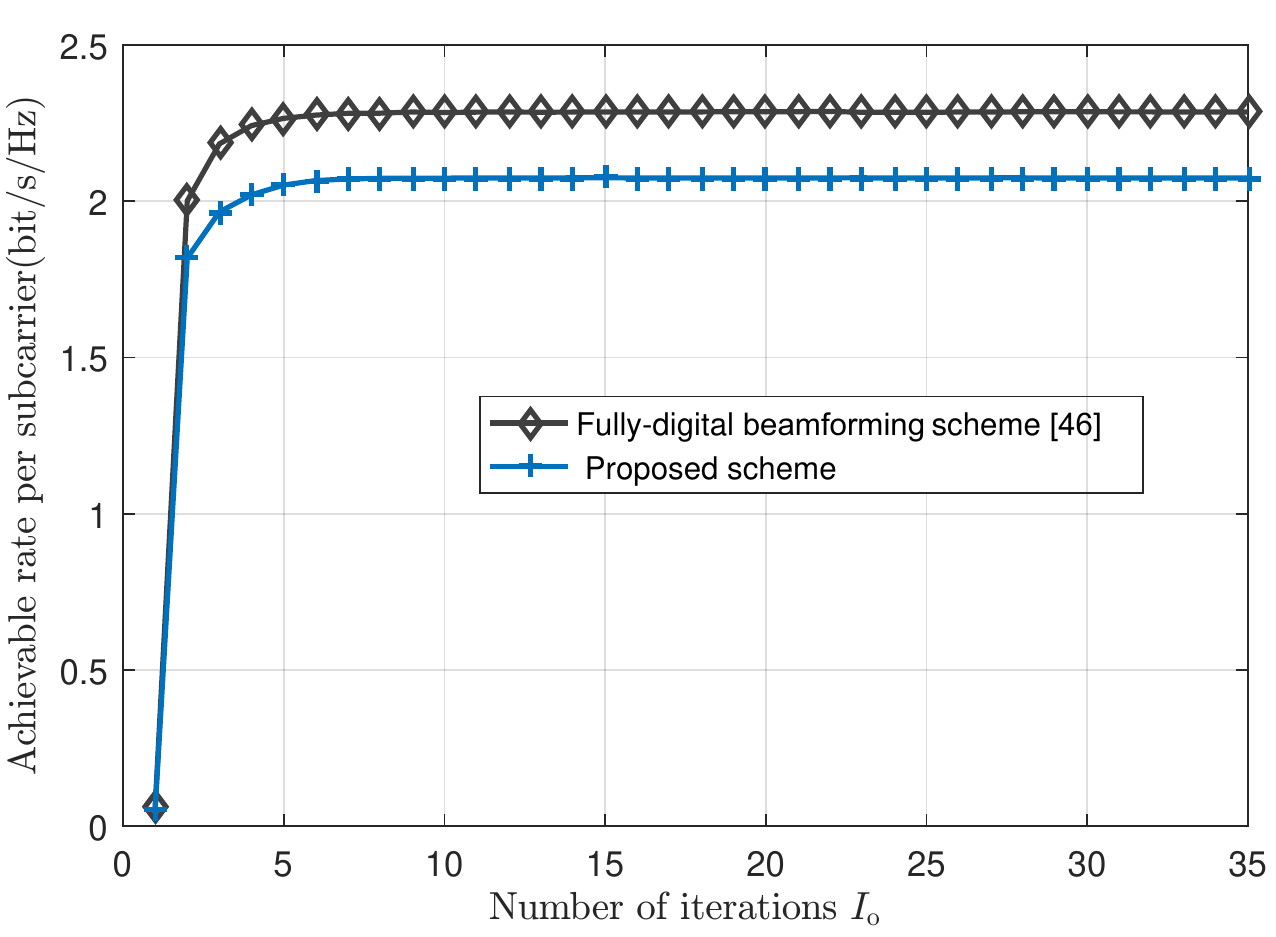}
	
	\caption{Achievable rate against the number of iterations $I_{\rm{o}}$, $N_{\rm{TX}}=16$.}
\end{figure}

To evaluate the convergence of the proposed scheme, Fig. 8 illustrates the achievable rate against the number of iterations $I_{\rm{o}}$. And we compare the performance between the proposed scheme and the optimal fully-digital beamforming design scheme~\cite{ref22} with $N_{\rm{TX}}=16$, $R=4$. One can observe that the achievable rate tends to stable after 5 iterations, which proves the effectiveness of the proposed scheme. Additionally, we can find that the achievable rate under fully-digital beamforming scheme is slightly higher than that of the proposed scheme, while the power consumption is huge for the former scheme due to the large number of RF chains.  Due to the small gap between them, it can be concluded that a near-optimal performance under the proposed scheme can be obtained.
\begin{figure}[htbp]
\centering
    \label{TD-PS} 
    \includegraphics[width=9cm,height=6cm]{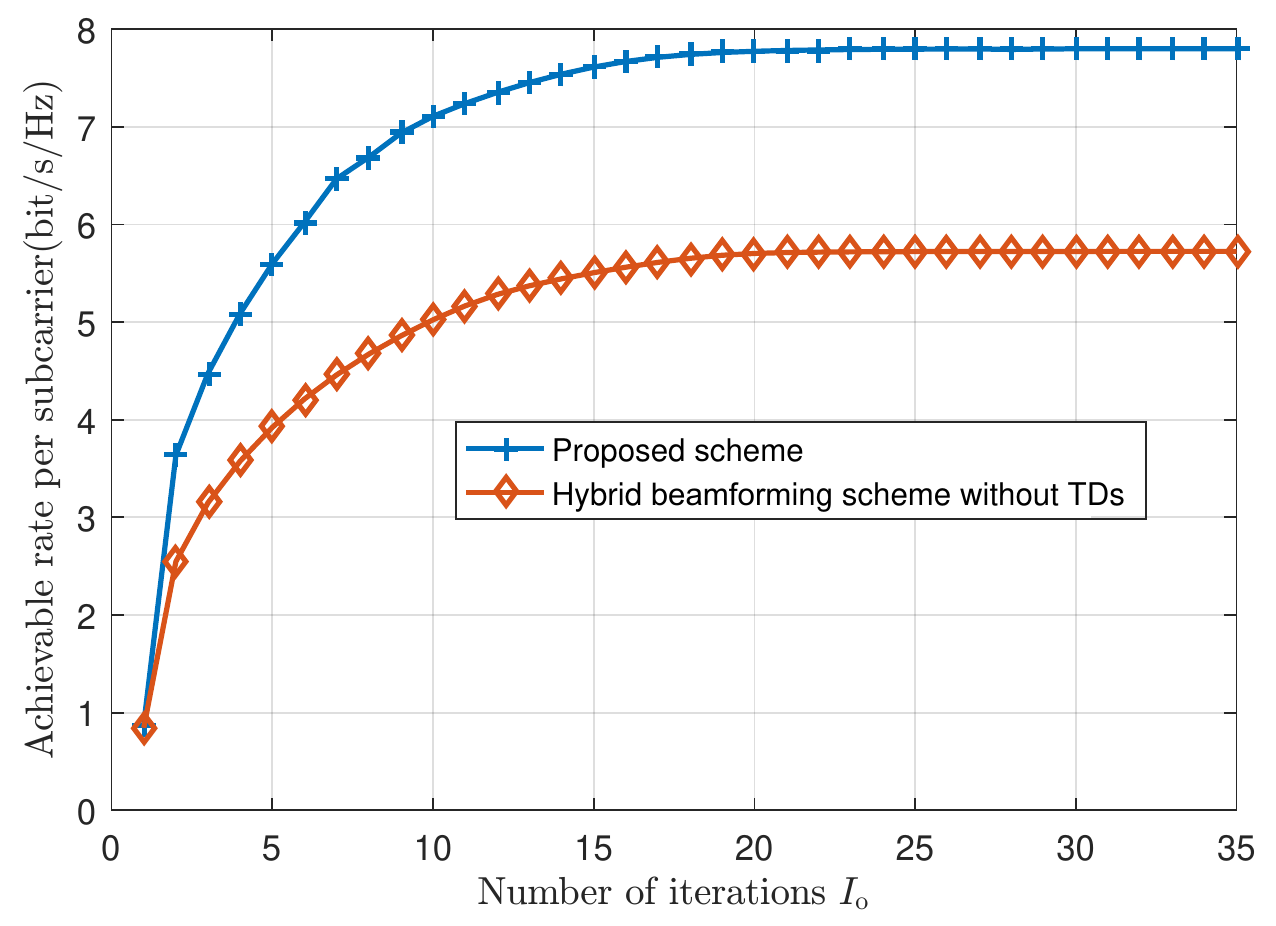}
	
	\caption{Achievable rate against the number of iterations $I_{\rm{o}}$, $N_{\rm{TX}}=256$.}
\end{figure}
\begin{figure}[htbp]
\centering
    \label{KT2subcarrier} 
    \includegraphics[width=9cm,height=6cm]{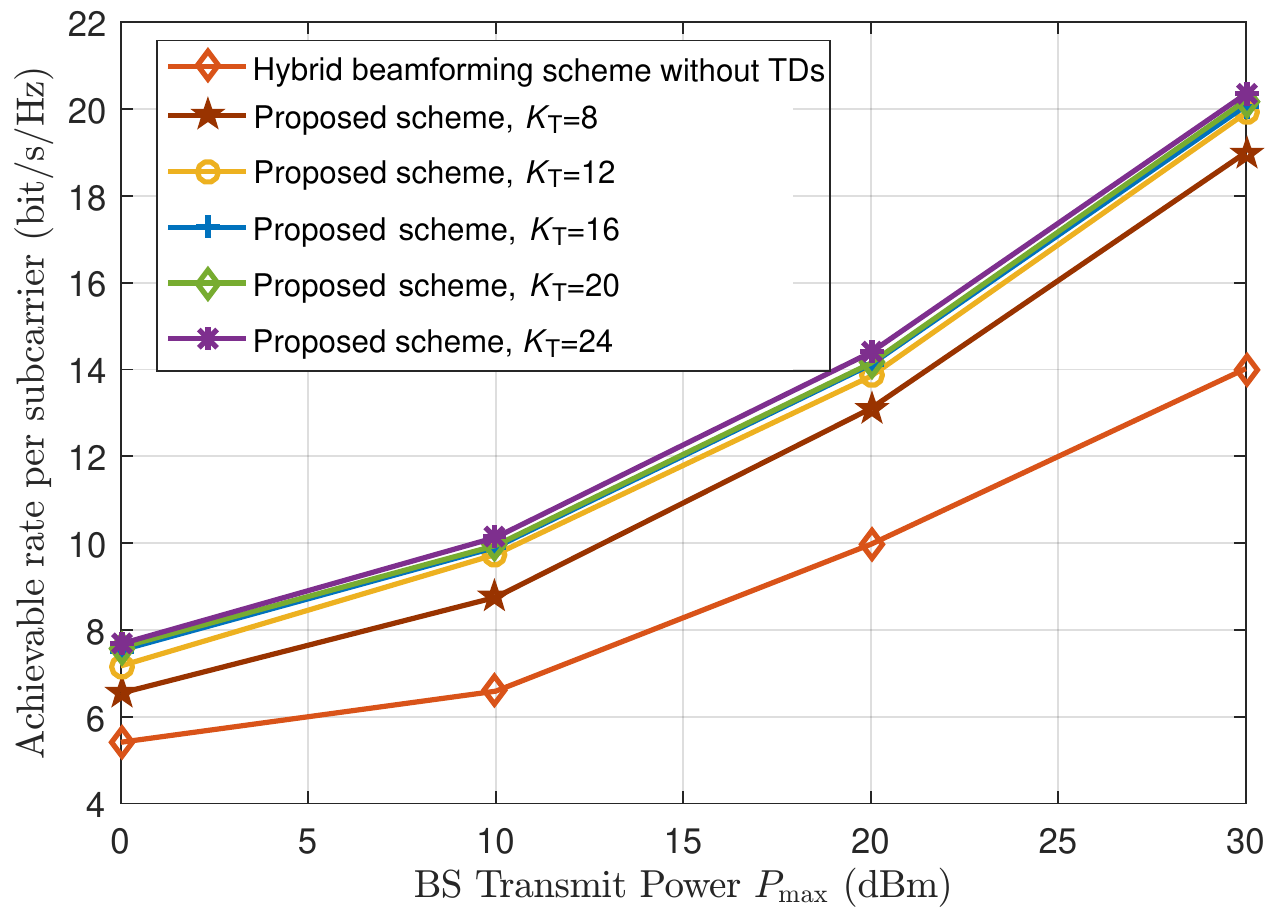}
	
	\caption{Achievable rate against maximum transmit power with different number of TDs $K_{\rm{T}}$.}
\end{figure}

In Fig.~9, we add the number of antennas to $N_{\rm{TX}}=256$ and compare the achievable rate under different schemes, where ``Hybrid precoding scheme without TDs" means that the analog beamforming is directly designed via (33). It is obvious that the achievable rate tends to stable after about 15 iterations. Compared this with Fig.~8, we can find that more antennas lead to more iterations for convergence. Additionally, one can observe that the achievable rate under the proposed scheme is higher than that of the hybrid precoding scheme without TDs. The main reason is that the TDs can make multiple beamforming focusing on RISs, such that RISs can reflect all signals to users, improving the SINRs of all the users.

Fig. 10 plots achievable rate versus maximum transmit power $P_{\rm {max}}$ with different number of TDs $K_{\rm T}$. One can easily find that the achievable rate increases with $P_{\rm{max}}$ under different schemes. Additionally, we find that more TDs can bring higher achievable rate, while the rate gap is smaller when $K_{\rm{T}}\geq 16$. This means that there is no necessary to deploy lots of TDs, and a near-optimal performance can be obtained with a few TDs, as the analysis results in~\cite{ref23}. On the other hand, more TDs result in the higher power consumption and hardware complexity. Thus, we need to tradeoff the system performance and cost so as to select appropriate number of TDs. Finally, it can be found that the achievable rate under hybrid precoding scheme without TDs is the lowest, and this is easy to understand.
\begin{figure}[htbp]
\centering
    \label{RIS_subcarrier} 
    \includegraphics[width=7cm,height=5cm]{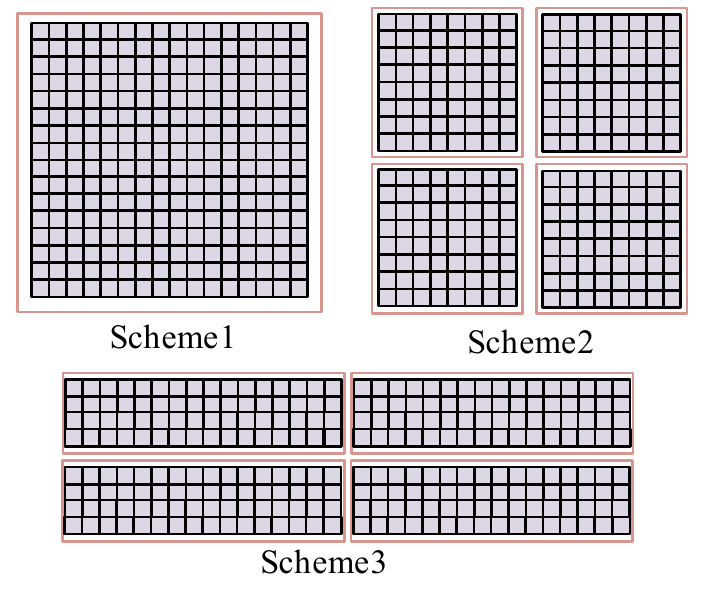}
	
	\caption{Different RIS deployment schemes.}
\end{figure}
\begin{figure}[htbp]
\centering
    \label{RISrectangular} 
    \includegraphics[width=9cm,height=6cm]{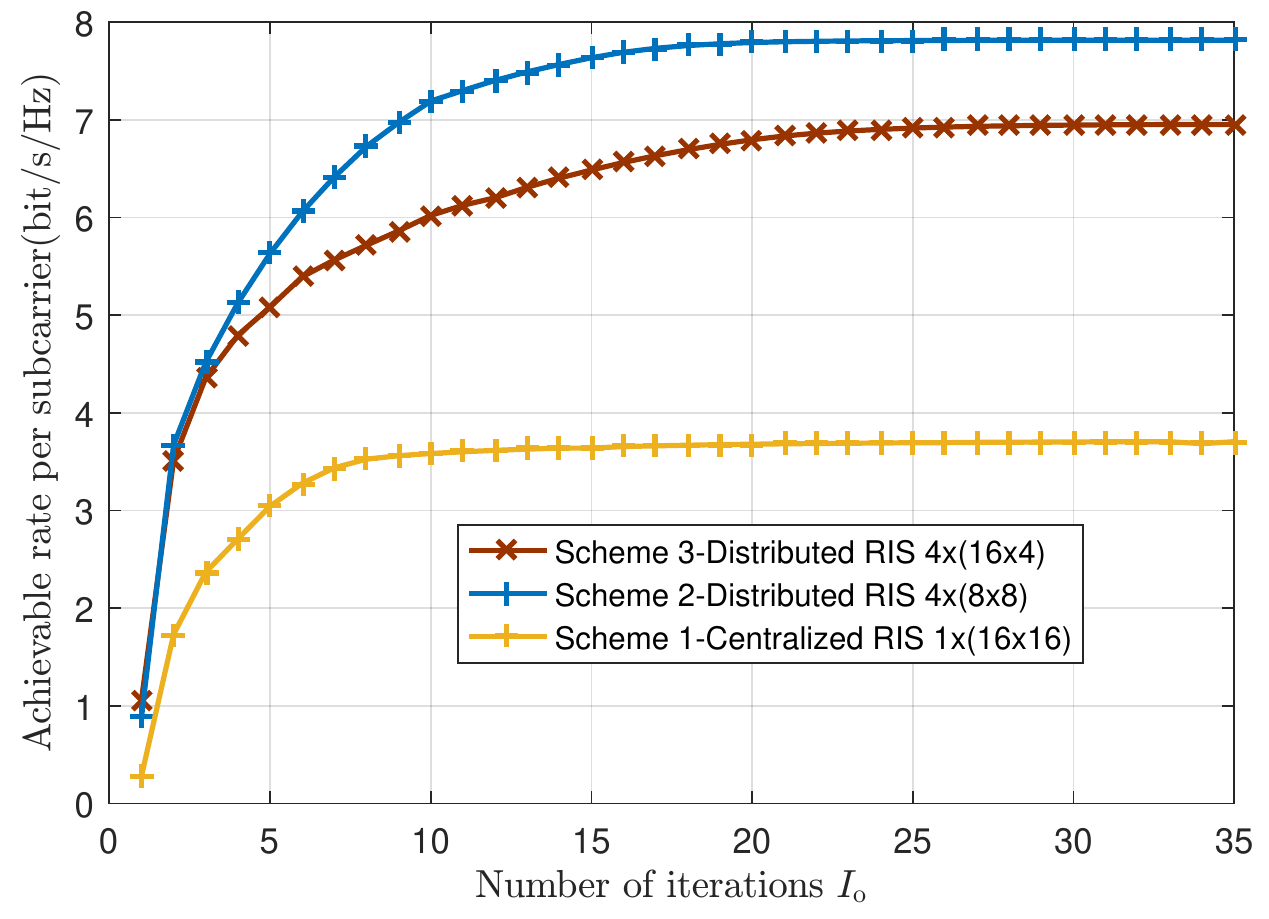}
	
	\caption{Achievable rate against the number of iterations $I_{\rm{o}}$ under different RIS deployment scheme.}
\end{figure}
\begin{figure}[htbp]
\centering
    \label{Bsubcarrier} 
    \includegraphics[width=9cm,height=6cm]{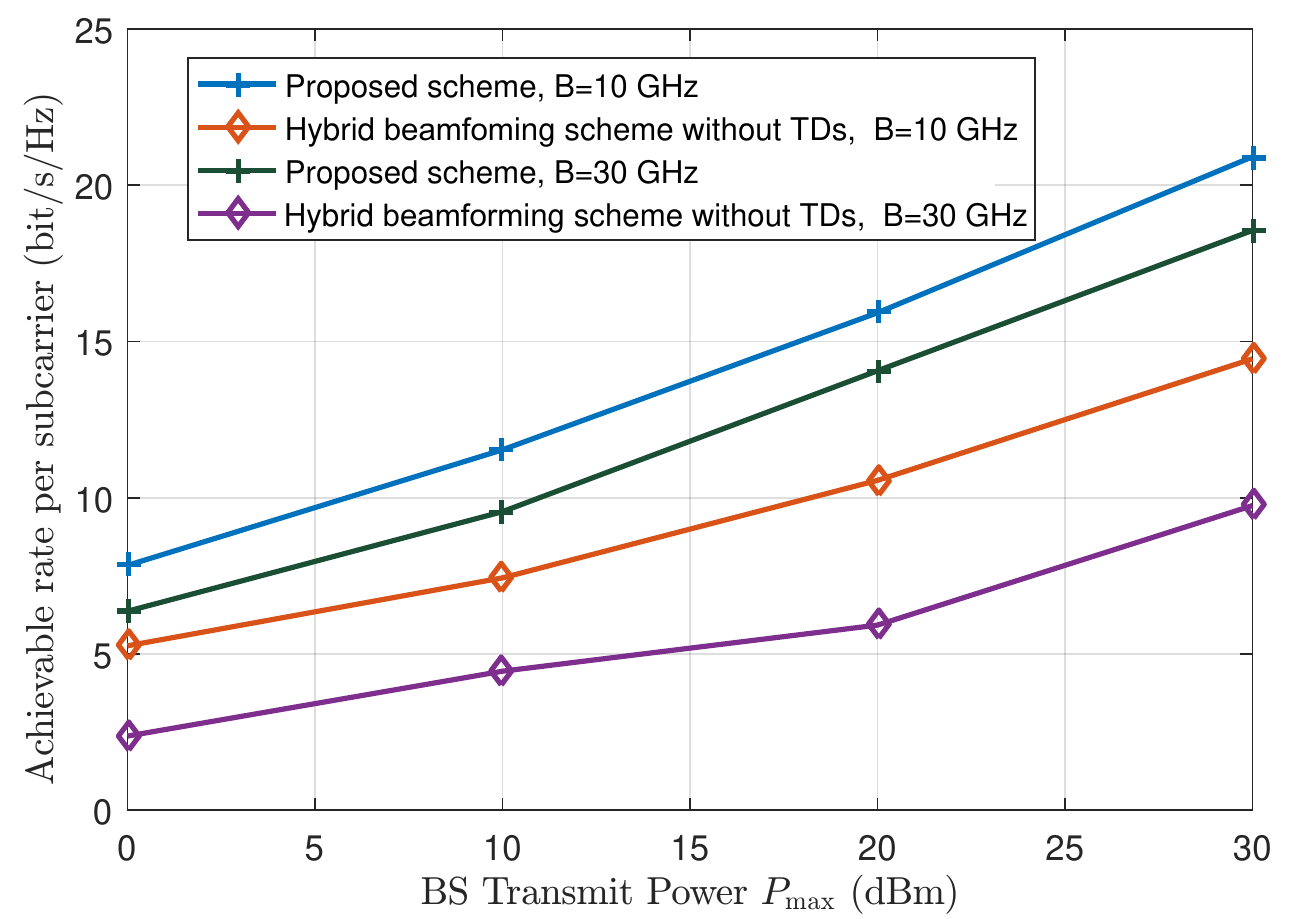}
	
	\caption{Achievable rate versus maximum transmit power under different bandwidth $B$.}
\end{figure}

To compare the achievable rate between the distributed RISs deployment and centralized RIS deployment, we consider three deployment schemes with total of 256 elements, and each is shown in Fig.~11. Scheme 1 is a $16\times 16$ centralized RIS deployment. Schemes 2 and 3 all includes 4 distributed RISs, their difference is that each one in Scheme 2 is a $8\times 8$ square RIS, while each one in Scheme 3 is a $16\times 4$ rectangular RIS. Schemes 2 and 3 are used to compare the beam split effect under different RIS shapes. On this basis, we plot Fig.~11 under different schemes. By analyzing Fig.~12, we can find three points based on the same condition, i)  the achievable rate with distributed RIS deployment (Scheme 2 or Scheme 3) is higher than that with centralized RIS deployment (Scheme 1), and the results are consistent with the analysis in Sec. II-B, namely the beam split effect with distributed RIS deployment is lower.  ii) the achievable rate with square RIS deployment (Scheme 2) is higher than that with rectangular RIS deployment (Scheme 3). iii) more iterations are needed for convergence under distributed RIS deployment.

Fig.~13 illustrates the achievable rate versus maximum transmit power $P_{\rm{max}}$ under different bandwidth $B$, where we set $N_{\rm{TX}}=256$ and $K_{\rm{T}}=16$. It can be easily found that the achievable rate increases with $P_{\rm {max}}$ under all schemes. Additionally, one can observe that the wider bandwidth leads to lower achievable rate with the same scheme. This is because that the beam split effect is more serious when the signal bandwidth is larger, as our analysis in Sec. II-B, which results in a low SINR and achievable rate. On the other hand, we can still find that the achievable rate with the proposed scheme is higher than that with conventional scheme.
\section{Conclusions}
In this paper, we considered a wideband THz RIS communication system, and investigated the normalized array gain and the beam split effect under different RIS sizes, shapes and deployments. Our research results show  that the beam split effect with square RIS deployment is smaller than that with rectangular RIS deployment under given conditions, meanwhile the beam split effect with distributed RIS deployment is smaller than that of the centralized RIS deployment. Then, we applied the FC-TD-PS-HB architecture at the BS and deployed the distributed square RISs to cooperatively mitigate the beam split effect by jointly designing hybrid analog/digital beamforming, time delays at the BS and the reflection coefficients at the RIS. Simulation results have verified that the proposed scheme can effectively relieve the beam split effect and improve system performance. In our future work, we will introduce TDs to RIS elements, and research how to reduce the beam split effect.

\end{document}